\documentclass[aps,prd,twocolumn,showpacs,floats,floatfix,letterpaper,nofootinbib,superscriptaddress,notitlepage]{revtex4-1}
\pdfoutput=1
\usepackage{hyperref}
\usepackage{amssymb,amsmath,latexsym,mathrsfs}
\usepackage{graphicx,subfigure}
\usepackage{epsfig}
\usepackage{varioref,xr-hyper}
\usepackage{color}
\usepackage{multirow}
\usepackage{array}
\usepackage{placeins}
\hypersetup{
     colorlinks   = true,
     citecolor    = green,
     linkcolor = cyan
     }
     
\newcommand\Tstrut{\rule{0pt}{2.6ex}}

\begin{document}

\title{Light path averages in spacetimes with non-vanishing average spatial curvature}
\author{S. M. Koksbang}
\email{sofie.koksbang@helsinki.fi}
\affiliation{Department of Physics, University of Helsinki and Helsinki Institute of Physics, P.O. Box 64, FIN-00014 University of Helsinki, Finland}

\begin{abstract}
Effects of inhomogeneities on observations have been vastly studied using both perturbative methods, N-body simulations and Swiss cheese  solutions to the Einstein equations. In nearly all cases, such studied setups assume vanishing spatial background curvature. While a spatially flat Friedmann-Lemaitre-Robertson-Walker model is in accordance with observations, a non-vanishing curvature is not ruled out. It is therefore important to note that, as has been pointed out in the literature, 1 dimensional averages might not converge to volume averages in non-Euclidean space. If this is indeed the case, it will affect the interpretation of observations in spacetimes with non-vanishing average spatial curvature. This possibility is therefore studied here by computing the integrated expansion rate and shear, the accumulated density contrast, and fluctuations in the redshift-distance relation in Swiss cheese models with different background curvatures. It is found that differences in mean and dispersion of these quantities in the different models are small and naturally attributable to differences in background expansion rate and density contrasts. Thus, the study does not yield an indication that the relationship between 1 dimensional spatial averages and volume averages depends significantly on background curvature.
\end{abstract}

\maketitle

\section{introduction}
The effects of inhomogeneities on light propagation are of fundamental importance in cosmology as these form the foundation for several useful observables including e.g. cosmic shear and CMB temperature fluctuations. The presence of structures in the Universe also represents a nuisance though, as inhomogeneities can lead to biases in mean values of observables, i.e. the mean of an observable might not converge to that expected based on the background Friedmann-Lemaitre-Robertson-Walker (FLRW) spacetime. (Except when otherwise explicitly mentioned, cosmic backreaction \cite{fluidI,fluidII,bc_review1,bc_review2} is not considered in the work presented here and accordingly a well-defined FLRW background is assumed to exist.) If such biases are not taken into account when interpreting observations, it can severely compromise parameter determinations in an era of precision cosmology and e.g. hinder a correct identification of dark energy parameters (see e.g. \cite{distanceCMB} for an example).
\newline\indent
Biased means are especially relevant for observations based on thin beams such as supernova observations where there is a significant risk that light rays do not trace spacetime fairly i.e. that the average expansion rate and density along light rays deviate significantly from their spatial averages \cite{misinterp,Linder_angle}. Such situations have been studied with exact solutions to the Einstein equations where light rays are not permitted to sample regions of density above a certain threshold \cite{postNewtonian, SC_vs_DR, SC_FleuryHelen} (see e.g. also \cite{stochastic}). If light rays on the other hand {\em are} permitted to trace spacetime fairly, only small biases in the mean of observables are expected. In particular, when using different types of averages such as area, angular, source and ensemble averages, certain observables will be unbiased in their mean while observables depending non-linearly on these will have small biases. This has been shown with studies based on perturbative expansions \cite{Kaiser_Peacock, source_average, lightcone,do_we_care, angle_vs_ensemble, angle_vs_ensemble_early, effect_DE} and is consistent with findings based on Swiss cheese models and similar \cite{syksyCMB,randomize,Ishak,tetris,tetris2,ltb_effect_hubble,cmb_digselv, postNewtonian} and on N-body simulations \cite{NbodyBolejko,misinterp}.
\newline\newline
Studies of effects of inhomogeneities on observations are usually conducted in settings where the average spatial curvature vanishes. This is reasonable since observations are generally consistent with vanishing curvature \cite{Planck18}. However, as recently argued in \cite{Curvature_serious}, the possibility of a small non-vanishing curvature should be considered since it is not excluded by observations. The currently existing literature on the topic of effects of inhomogeneities on light propagation should therefore be complemented with studies that take the possibility of non-vanishing average spatial curvature into account. Specifically, as pointed out in \cite{Tardis}, it is not clear that 1 dimensional spatial averages converge to volume averages in curved space because 1 dimensional line integrals and volume integrals have different measures when space is not Euclidean. This is important for observations since averages along light rays are related to 1 dimensional spatial averages if structures evolve slowly compared to the time it takes a light ray to traverse the homogeneity scale (assuming statistical homogeneity and isotropy); if these conditions are fulfilled, the time evolution can be neglected during such a time interval, and hence the light ray average over such a distance simply becomes a 1 dimensional spatial average (see also \cite{Tardis} and e.g. \cite{av_obs1} for detailed considerations on similar matters). Thus, if 1 dimensional averages do not converge to volume averages in curved space or if the convergence is much slower than in flat space, then biases of mean values of observations are possibly much larger and the dispersion around the mean may be more significant than in the flat (Euclidean) case.  This could e.g. be important for studies of mean and dispersion in the Hubble diagram which have been conducted e.g. with focus on the $H_0$-problem \cite{source_average, H0_lightcone, Io1,Io2,Io3, TurnerH0, WojtakH0}. Effects of non-vanishing curvature may indeed be especially important for low-redshift observations since studies indicate that cosmic backreaction may lead to the emergence of spatial curvature at late times (see e.g. \cite{emergence}).
\newline\newline
The purpose of the presented work is to study if and how a non-vanishing spatial background curvature affects the relationship between volume averages and light path averages with special emphasis on the effects on the mean and dispersion of observables. This is done by considering light propagation in Swiss cheese models with Lemaitre-Tolman-Bondi (LTB) structures \cite{LTB1,LTB2,LTB3} placed in curved FLRW backgrounds. Specifically, the mean and dispersion of the redshift-distance relation will be sampled together with light path averages of the density contrast and the integrated expansion rate and shear in spacetimes with different curvature.
\newline\newline
The paper is organized as follows. Section \ref{sec:modelSetup} describes the models used for the study while section \ref{sec:lightpropagation} gives a brief review of light propagation. Results are presented and discussed in section \ref{sec:Results} while section \ref{sec:Summary} provides a summary.

\section{Model setup}\label{sec:modelSetup}
\begin{figure}
\centering
\includegraphics[scale = 0.5]{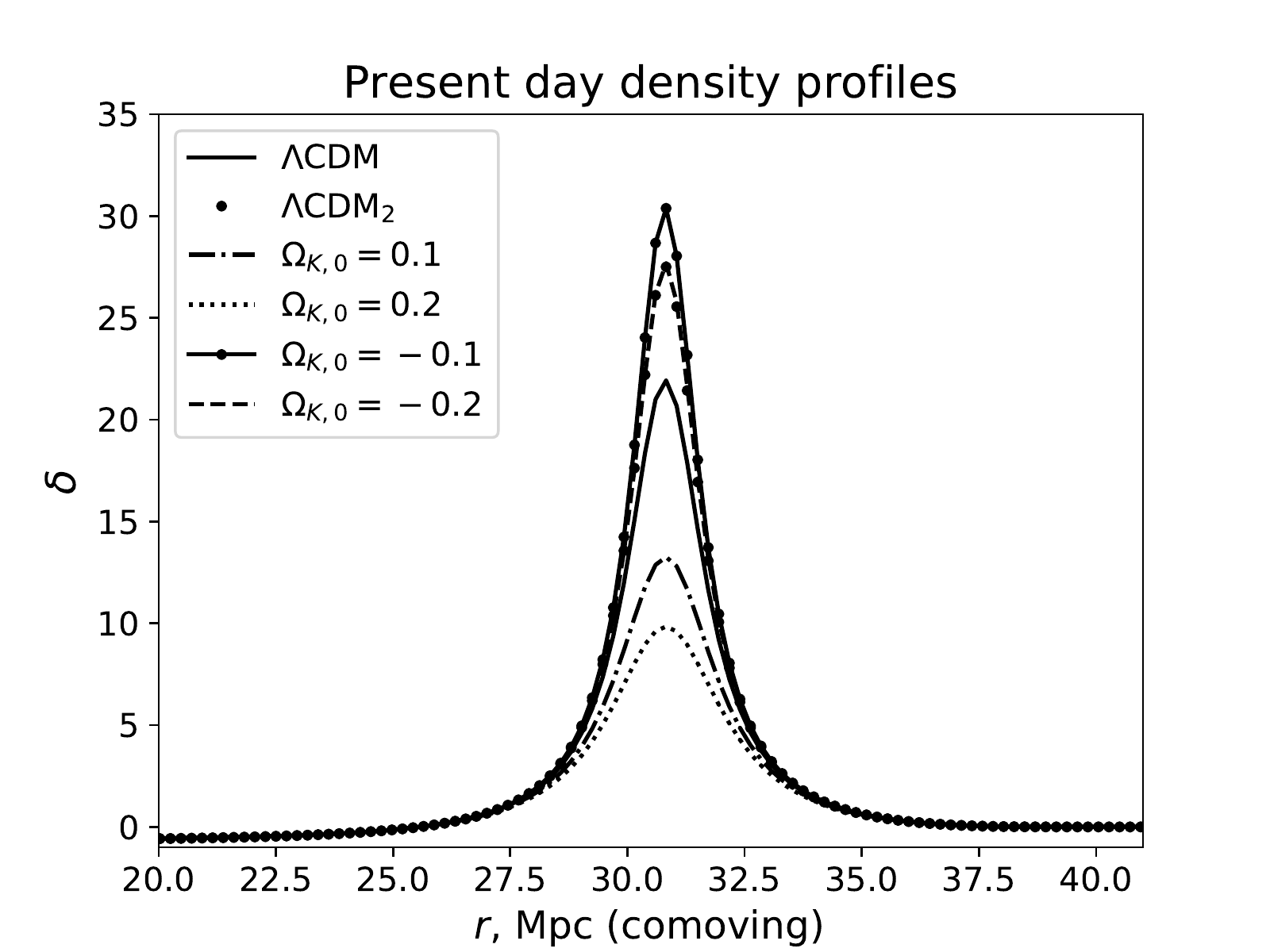}
\caption{Present time 1D density profiles of LTB structures. Models are named according to table \ref{table:models}. The density profiles cannot be distinguished at small values of $r$ and are therefor not shown for $r<20$Mpc. Note that the density profiles of the models $\Lambda$CDM$_2$ and $\Omega_{K,0} = -0.2$ lie almost exactly on top of each other. Their combined graph therefore looks very similar to that of the model with $\Omega_{K,0} = -0.1$ which, however, has a slightly larger overdensity.}
\label{fig:LTBdensity}
\end{figure}
\begin{figure}
\centering
\includegraphics[scale = 0.5]{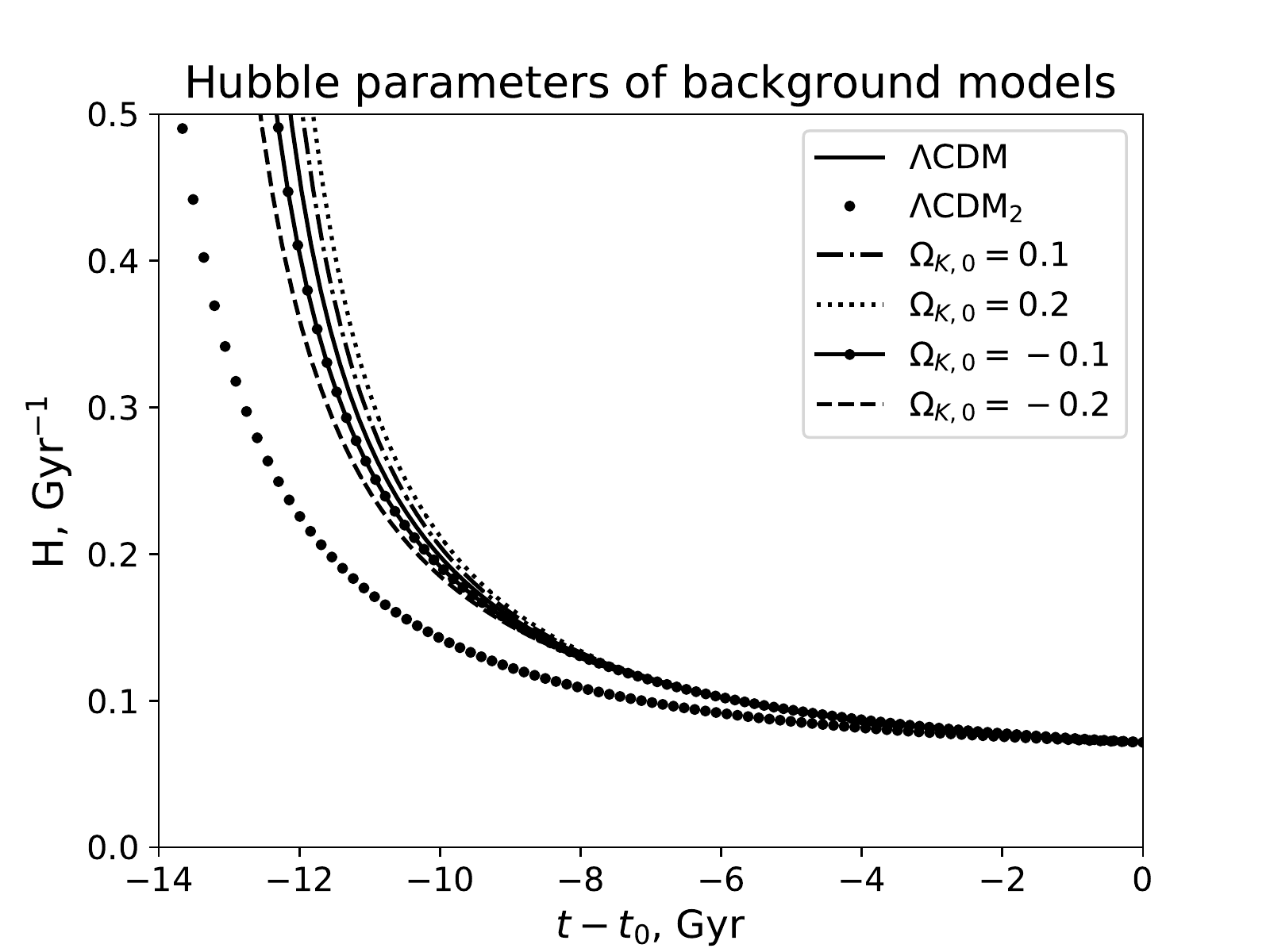}
\caption{Hubble parameters of backgrounds models. Models are named according to table \ref{table:models}.}
\label{fig:bgH}
\end{figure}
\begin{figure}
\centering
\includegraphics[scale = 0.5]{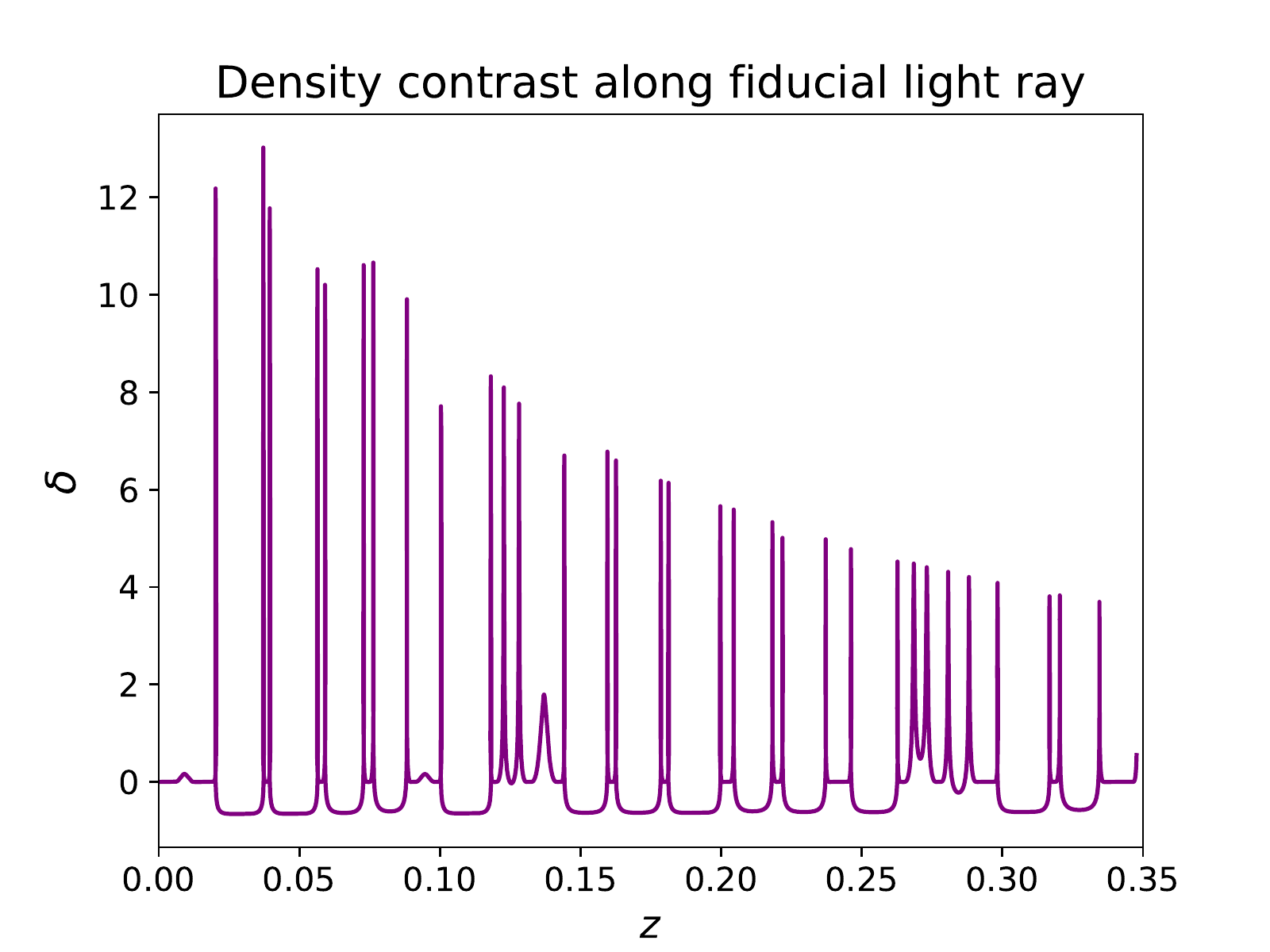}
\caption{Density contrast along fiducial light ray in the Swiss cheese model with a $\Lambda$CDM background.}
\label{fig:rho_along_lr}
\end{figure}
Swiss cheese models will be constructed to imitate statistically homogeneous and isotropic universes with inhomogeneities modeled as mass-compensated voids mimicking the large scale structures of the Universe. The structures are described using the spherically symmetric LTB metric corresponding to the line element
\begin{equation}
ds^{2} = -c^2dt^2 +\frac{A^2_{,r}(t,r)}{1-k(r)}dr^2 +A(t,r)^2d\Omega^2.
\end{equation}
Subscripted commas followed by a coordinate indicate partial derivatives.
\newline\newline
The time dependence of the LTB spacetimes is dictated by $A$ which has an evolution determined by
\begin{equation}\label{eq:A}
\frac{1}{c^2}A_{,t}^2 = \frac{2M}{A} - k +\frac{1}{3c^2}\Lambda A^2.
\end{equation}
The inhomogeneous dust density is given by
\begin{equation}
\rho = c^4\frac{M_{,r} }{4\pi G_N A^2A_{,r} }.
\end{equation}
$M(r)$ is an integration constant and $G_N$ is Newton's constant.
\newline\newline
LTB models can be used to model central voids surrounded by mass-compensating overdensities which are again surrounded by a homogeneous FLRW spacetime referred to as the background. When the big bang occurs at the same value of the $t$-coordinate everywhere in space, the structures will have no decaying modes \cite{decaying} and the entire LTB space will tend towards the FLRW background at early times. The models used here are all specified by this attribute and are further specified by $A(t_{1200},r) = \frac{1}{1200}r$, where $t_{1200}$ is the time at which the background scale factor reaches $\frac{1}{1200}$. LTB models are covariant under transformations of the radial coordinate and specifying $A(t_{1200},r)$ as above corresponds to fixing $r$. Setting $A(t_{1200},r) = \frac{1}{1200}r$ is a convenient way to ensure (for conceptual convenience) that $A = ar$ in spacetime regions that are FLRW i.e. in the region outside the LTB structure as well as the entire LTB space at early times (asymptotically). The time $t_{1200}$ is chosen because setting initial conditions at early times was found to give faster and more numerically stable results for the specific models studied here. It should be noted though, that LTB models are generally not appropriate for cosmological studies at early times where radiation is non-negligible. However, there is no reason to expect that neglecting radiation when setting initial conditions will significantly affect the current study where the models are only considered at late times. See e.g. \cite{pressure1,pressure2} for considerations of how pressure affects structure formation.
\newline\indent
The final specification of the inhomogeneity profile is determined by setting 
\begin{equation}\label{eq:k}
k = \left\{ \begin{array}{rl}
-k_{max}\cdot 10^{-8}r^2\left(\left(\frac{r}{r_b} \right)^{p_1} -1 \right)^{p_2} +k_{bg}r^2  &\text{if} \,\, r\leq r_b \\
k_{bg}r^2 &\text{if} \,\, r>r_b
\end{array} \right.,
\end{equation}
with $r_b = 40$Mpc the comoving radius of the inhomogeneity and $k_{max}$ a constant. The parameters $p_1$ and $p_2$ are discussed further below. The constant $k_{bg}$ is the curvature parameter of the background i.e. $k_{bg} = \frac{\pm1}{R_0^2}=\frac{H_0^2}{c^2}\left( \Omega_{tot,0}-1\right) $, where $R_0$ is the background curvature radius and $\Omega_{tot,0}$ the total present time density parameter. Different backgrounds are considered. These are specified by $H_0 = 70$km/s/Mpc and the density parameters given in table \ref{table:models}. The specific values of the density parameters and hence of the background curvature parameter, $k_{bg}$, were chosen based on two competing considerations: On one hand, to ensure the detection of effects of background curvature, the curvature parameters should be chosen somewhat larger than what can be expected to possibly be relevant for the real Universe. On the other hand, the dynamics of the background as well as the dynamics and size of the inhomogeneities should be as similar as possible in all the considered models to ensure that detected differences are in fact due to curvature and not, say, differences in background expansion rates. In relation to the latter consideration, i.e. in order to quantify effects from modifying the dynamics of the models, two different flat models are considered.
\newline\indent
Table \ref{table:models} shows the values of $k_{max}$ used in each model. The specific values of $k_{max}$ were chosen based on the following considerations: First of all, the values should be similar within each model in order for the models to differ as little as possible aside from their background curvature but, second of all, the values should not be chosen so small that the density fluctuations are linear. The latter implies that $k_{max}$ cannot be chosen to be the same for all models as the minimum $k_{max}$ necessary to obtain non-linear density fluctuations in the models with largest background densities is larger than the maximum possible value of $k_{max}$ that does not lead to shell crossings in the outer layers of the LTB inhomogeneities in the models with the smallest background densities. Based on these considerations, $k_{max}$ is chosen to be equal to $5.4$Mpc$^{-2}$ for most models since this value is just large enough to get the structures of model $\Omega_{K,0}=-0.2$ into the clearly non-linear regime at present times (with a maximum density contrast of $\delta \sim10$). For the two models with $\Omega_{K,0}>0$, $k_{max}$ has to be chosen slightly smaller, namely as $5.3$Mpc$^{-2}$ and $5.1$Mpc$^{-2}$ while it has to be chosen as small as $4$Mpc$^{-2}$ for the model $\Lambda$CDM$_2$ in order to avoid shell crossings at or before present time.
\newline\indent
Regarding the parameter choices for $k(r)$, note that the details of the resulting LTB structure including e.g. the density profile does not seem to have great affect on light propagation over long distances at a statistical level. This is e.g. illustrated in \cite{cmb_digselv} where the distance to the CMB is studied using four different Swiss cheese models based on two LTB and two Szekeres (\cite{Szekeres}) models. All four models of \cite{cmb_digselv} have $k(r)$ similar to that used here but with different values of $p_1$ and $p_2$. The study of \cite{cmb_digselv} specifically indicates that the values chosen for $p_1$ and $p_2$ only affect results where high precision is very important. For instance, choosing $p_1 = 2$ and $p_2 = 4$ or $p_1 = p_2  = 6$ both leads to distributions in fluctuations in the angular diameter distances to the CMB to cover intervals of order 0.01-0.1. On the other hand, the mean values obtained from these models are distributed in the interval $10^{-4}-10^{-6}$, a difference which could be important for high-precision studies. Here, it is more interesting to note that the present day density profile of the resulting LTB structure becomes more bucket-shaped when $p_1$ and $p_2$ increase and more cone-shaped when they decrease. At the same time, larger values of $p_1$ and $p_2$ makes it more time consuming to numerically solve the ODEs describing light propagation. Larger values of $p_1$ and $p_2$ also make the mass-compensating overdensities surrounding the voids more prominent. Here, $p_1 = p_2 = 6$ was chosen to give somewhat bucket-shaped central voids surrounded by clearly non-linear overdensities without increasing computation time significantly compared to choices with smaller values of $p_1$ and $p_2$.
\newline\indent
When choosing the parameters for $k(r)$, $r_b$ is of central importance as it determines the size of the LTB structure. As was also mentioned in \cite{cmb_digselv}, choosing $r_b = 40$Mpc is similar to the choices made in Swiss cheese studies by other authors such as \cite{syksyCMB,Ishak} and leads to present day void radii of approximately $38$Mpc which is in agreement with observations \cite{voids_2001,voids_2003,Voids_2002}, albeit to the slightly larger side. Note, however, that the observed size, volume fraction and deepness of real voids depends much on the involved void definition. Note also that the choice of $r_b$ determines the homogeneity scale of the Swiss cheese model since the considered LTB models reduce {\em exactly} to their FLRW backgrounds at this point. With $r_b = 40$Mpc, the present time homogeneity scale of the model is 80Mpc which is close to the believed homogeneity scale of the real universe of order of $100$Mpc. The reason a somewhat small homogeneity scale is chosen here is that the homogeneity scale is set by a single structure while it in the real universe is determined by a complex network of structures on many scales. Requiring a larger homogeneity scale for the considered Swiss cheese models would thus e.g. lead to introducing very large voids.
\newline\newline
The present time density profiles of the considered models and their background Hubble parameters are shown in figures \ref{fig:LTBdensity} and \ref{fig:bgH}. In addition, the density profile along a single light ray in the considered Swiss cheese model with an ordinary $\Lambda$CDM background is shown in figure \ref{fig:rho_along_lr}. The light ray is chosen randomly amongst the studied light rays in that particular Swiss cheese model (see the next section for computational details). For comparison, the reader may want to look at figure 3 in \cite{NbodyBolejko} which shows the density profiles along a light ray in the Millennium simulation \cite{millenium} and in different inhomogeneous models including two Swiss cheese models. The complexity in the density contrast along the light ray traced through the Millennium simulation specifically emphasizes the simplicity of Swiss cheese models based on only a single structure size. This simplicity may be problematic for some studies but for the current purpose it is not an issue; using a single structure size should be sufficient to determine if observations in an inhomogeneous universe are affected by background curvature at a statistical level.

\begin{table}[]
\centering
\begin{tabular}{c c c c}
\hline\hline
\Tstrut
Model & $\Omega_{m,0}$ & $\Omega_{\Lambda}$ & $k_{max}$ (Mpc$^{-2}$)\\
\hline
$\Lambda$CDM & 0.3 & 0.7 & 5.4\\
$\Lambda$CDM$_2$ & 0.2 & 0.8 & 4\\
$\Omega_{K,0} = 0.1$ & 0.35 & 0.75 & 5.4\\
$\Omega_{K,0} = 0.2$ & 0.4  & 0.8 & 5.4\\
$\Omega_{K,0}= -0.1$ & 0.25 & 0.65 & 5.3\\
$\Omega_{K,0}= -0.2$ & 0.2 & 0.6 & 5.1\\
\hline
\end{tabular}
\caption{Specification of backgrounds and $k_{max}$ used for the different Swiss cheese models considered. The models will be referred to by the background values of $\Omega_{K,0}:=\Omega_{m,0}+\Omega_{\Lambda}-1$.}
\label{table:models}
\end{table}

\subsection{Swiss cheese construction}\label{subsec:cheese}
The Swiss cheese models are constructed on the fly by turning light rays around when they reach $r = r_b+1$Mpc. The choice of placing the turnaround point at $r = r_b +1$Mpc is based on the desire to increase effects of inhomogeneity: Choosing the turnaround point to be close to the boundary $r_b$ leads to a high effective packing fraction of the structures. At each turnaround point, the light rays are directed towards the LTB inhomogeneity with a random impact parameter. The series of impact parameters for each light ray is saved and re-used for the other studied models. This is done because only 1000 light rays will be considered for each model. Although this should be large enough to obtain trustworthy results, there is a risk that sample variance will be non-negligible. By providing the same series of impact parameters for each model, the sample variance should be similar between the models. This should protect against false positives i.e. false identifications of effects of background curvature that are in reality just due to sample variance.
\newline\indent
An alternative to constructing Swiss cheese models on the fly is to construct a fixed Swiss cheese spacetime. This can e.g. be done using the Jodrey-Tory algorithm \cite{Jodrey,Jodrey2} (see e.g. \cite{JT_curved} for a modification for curved space). There are some disadvantages with this approach when working with curved backgrounds though. Major disadvantages are the memory consumption needed to store structure locations and the resources required to monitor the distances between a light path and the structures of the spacetime when propagating a light ray
\footnote{The maximum packing fraction for a random close packing of spheres is approximately 0.64 \cite{RCP_experimental}. Therefore if, for instance, an observer is permitted to look in any direction and required to see an inhomogeneous universe out to at least 1Gpc, a total number of approximately 19,100 LTB structures are necessary (with $r_b = 40$Mpc). Each LTB structure is defined through the comoving spatial coordinates of its center (three numbers of type double) and possibly an integer identification number. All in all, this corresponds to approximately 0.5 MB. This can be quite a lot for a laptop (and can e.g. typically not be allocated on the stack), especially when adding the fact that the same program has to save data computed along each light ray, but should be manageable when running on a cluster. However, the main difficulty with such a large number of structures is the computation time required to monitor the proximity of a light ray to structures during time periods when the light ray propagates in the cheese.}.
These problems can be remedied by considering a moderately sized spatial region with periodic boundary conditions as was done in \cite{cmb_digselv, dig_og_cc}. However, when dealing with a curved spacetime, choosing appropriate boundary conditions and pairing of sides becomes non-trivial (see e.g. \cite{boundary_conditions_hyperbolic} for a discussion and examples for the hyperbolic plane). When wishing to compare results based on differently curved spaces it is especially an issue how to pair sides within each model to obtain similar inhomogeneity distributions and identical topologies of the different models. If this cannot be done sensibly, there is a risk of increased effects of variance between the computed data sets from the different models. This is the main reason an on-the-fly construction of Swiss cheese models is used here.

\section{Light propagation}\label{sec:lightpropagation}
Light propagation in LTB models is well documented in the existing literature, including in some of the references to Swiss cheese studies already given. This section therefore only gives a very brief description of the specific method used here.
\newline\newline
The redshift is defined by $z:=\frac{\left( k^{\alpha}u_{\alpha}\right)_e}{\left( k^{\beta}u_{\beta}\right)_0}$, with the subscript $0$ implying evaluation at the point of observation and $e$ at the point of emission. The LTB models are given in a synchronous and comoving foliation so $u^{\alpha}=(1,0,0,0)$. The null tangent vector $k^{\alpha}$ can be found by solving the geodesic equations
\begin{equation}
\frac{d}{d\lambda}\left(g_{\alpha\beta}k^{\beta} \right) = \frac{1}{2}g_{\mu\gamma,\alpha}k^{\mu}k^{\gamma},
\end{equation}
where $\lambda$ is the affine parameter of the null geodesic and $g_{\alpha\beta}$ the metric tensor.
\newline\indent
The angular diameter distance will be sampled along the light rays together with the redshift. The angular diameter distance is computed by solving the transport equation
 \begin{equation}\label{D_dot}
\frac{ d^2D^a_b}{d\lambda^2} = T^a_cD^c_b.
\end{equation}
The components of $T_{ab}$ are given in terms of the Riemann tensor, $R_{\alpha\beta\mu\nu}$, the Ricci tensor, $R_{\mu\nu}$, and the vectors spanning the 2 dimensional Euclidean space orthogonal to the light path in the observer rest frame, $E_1^{\mu}, E_2^{\mu}$, combined as $\epsilon^{\mu} := E_1^{\mu} - iE_2^{\mu}$:  
\begin{equation}
T_{ab} = 
  \begin{pmatrix} \mathbf{R}- Re(\mathbf{F}) & Im(\mathbf{F}) \\ Im(\mathbf{F}) & \mathbf{R}+ Re(\mathbf{F})  \end{pmatrix},
\end{equation}
with $\mathbf{R}: = -\frac{1}{2}R_{\mu\nu}k^{\mu}k^{\nu}$ and $\mathbf{F}:=-\frac{1}{2}R_{\alpha\beta\mu\nu}(\epsilon)^{\alpha}k^{\beta}(\epsilon)^{\mu}k^{\nu}$.
\newline\newline
It is also interesting to note that the redshift in a general dust+$\Lambda$ spacetime with a comoving, synchronous spacetime foliation and well-defined FLRW background can be written as (see e.g. \cite{av_obs1})
\begin{equation}\label{eq:z}
\begin{split}
1+z &= e^{\int_{t(\lambda)}^{t_0}dt\left(\frac{1}{3}\Theta + c^2\sigma_{\alpha}^{\beta}e^{\alpha}e_{\beta} \right)  }\\ &= e^{\int_{t(\lambda)}^{t_0}dtH }\cdot e^{\int_{t(\lambda)}^{t_0}dt\left(\frac{1}{3}\Delta\Theta + c^2\sigma_{\alpha}^{\beta}e^{\alpha}e_{\beta} \right)  }\\
&=(1+z_{bg})\cdot e^{\int_{t(\lambda)}^{t_0}dt\left(\frac{1}{3}\Delta\Theta + c^2\sigma_{\alpha}^{\beta}e^{\alpha}e_{\beta} \right)  },
\end{split}
\end{equation}
where $H$ is the background Hubble parameter, $z_{bg}$ is the background redshift, $\Theta$ is the local expansion rate and $\sigma_{\alpha}^{\beta}e^{\alpha}e_{\beta}$ the shear projected onto the spatial direction of the light ray.
\newline\indent
According to this expression, the observed redshift will deviate from the background redshift if the integrals of the fluctuations in the expansion rate, $\Delta \Theta$, and of the projected shear do not cancel (individually or with each other) along the light ray. For specific LTB models it has been found to be the case that these two contributions cancel with each other to a high precision \cite{Tardis,cmb_digselv}. This cancellation has also been shown to occur for perturbed FLRW models \cite{nearFRW}. The cancellation was, however, not found to occur in the recently presented (somewhat exotic) LTB Swiss cheese model in \cite{din_preview}.

\section{Results}\label{sec:Results}
\begin{figure}
	\centering
	\subfigure[]{
		\includegraphics[scale = 0.5]{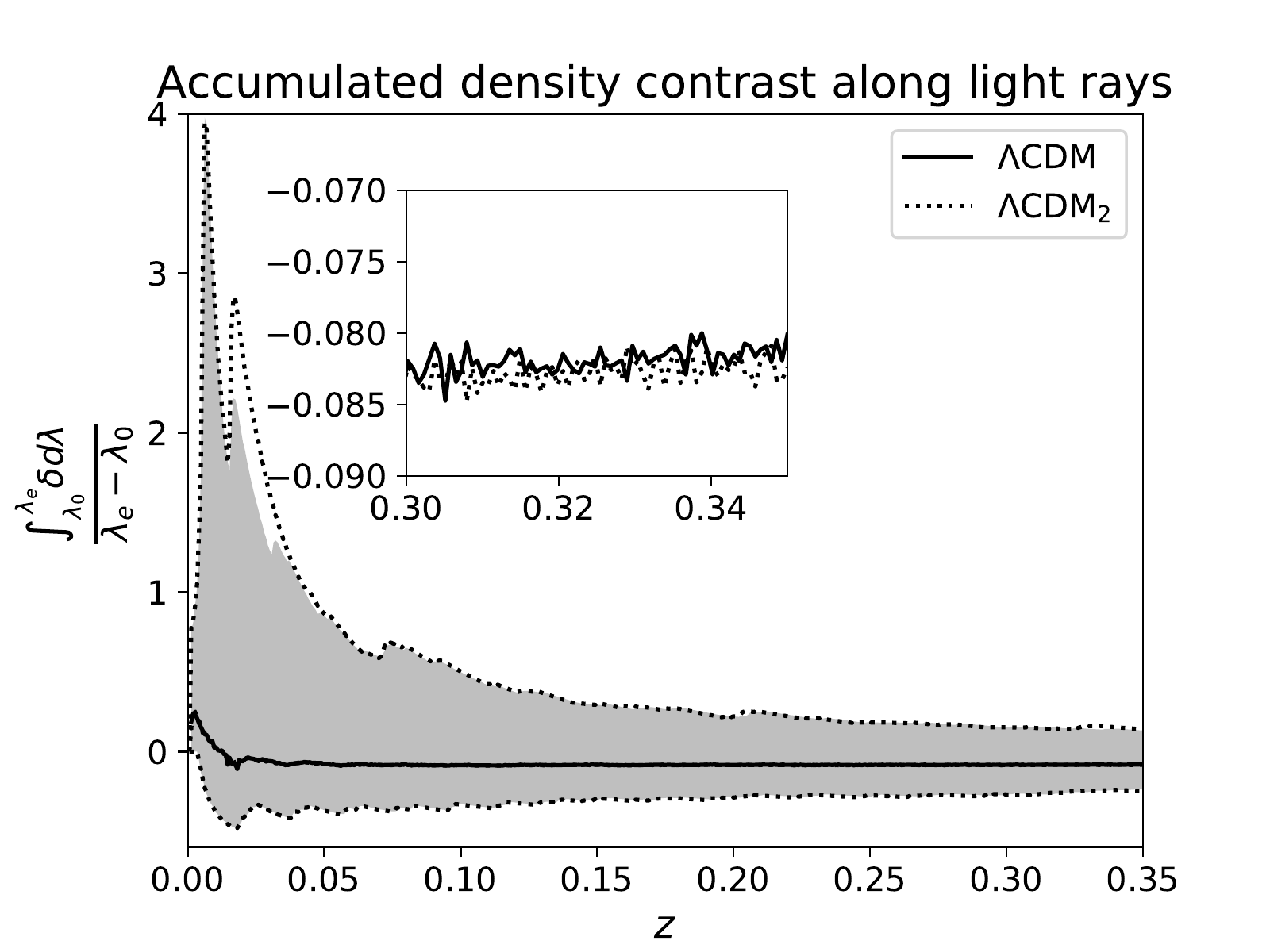}
	}\par
	\subfigure[]{
		\includegraphics[scale = 0.5]{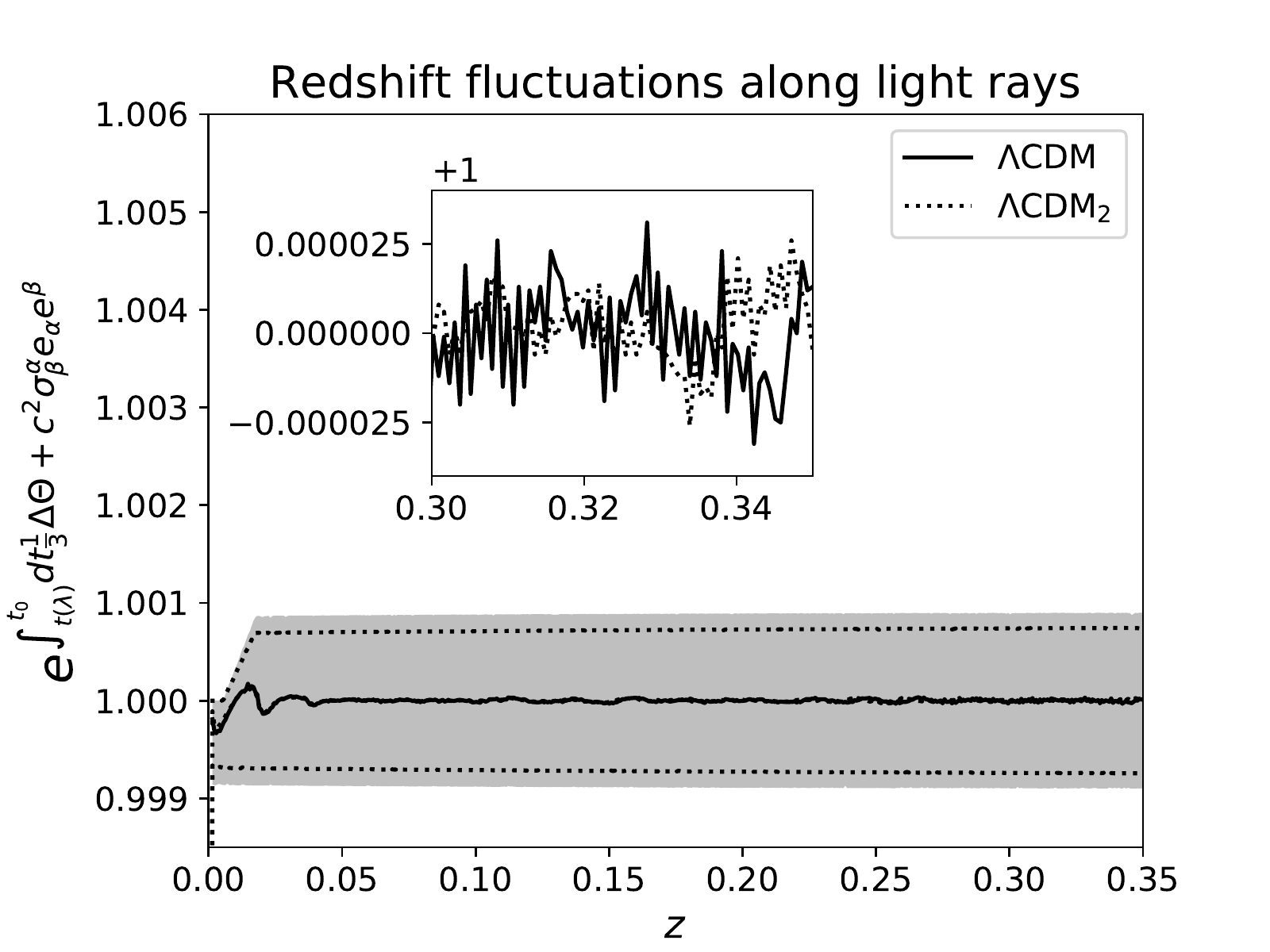}
	}\par
	\subfigure[]{
		\includegraphics[scale = 0.5]{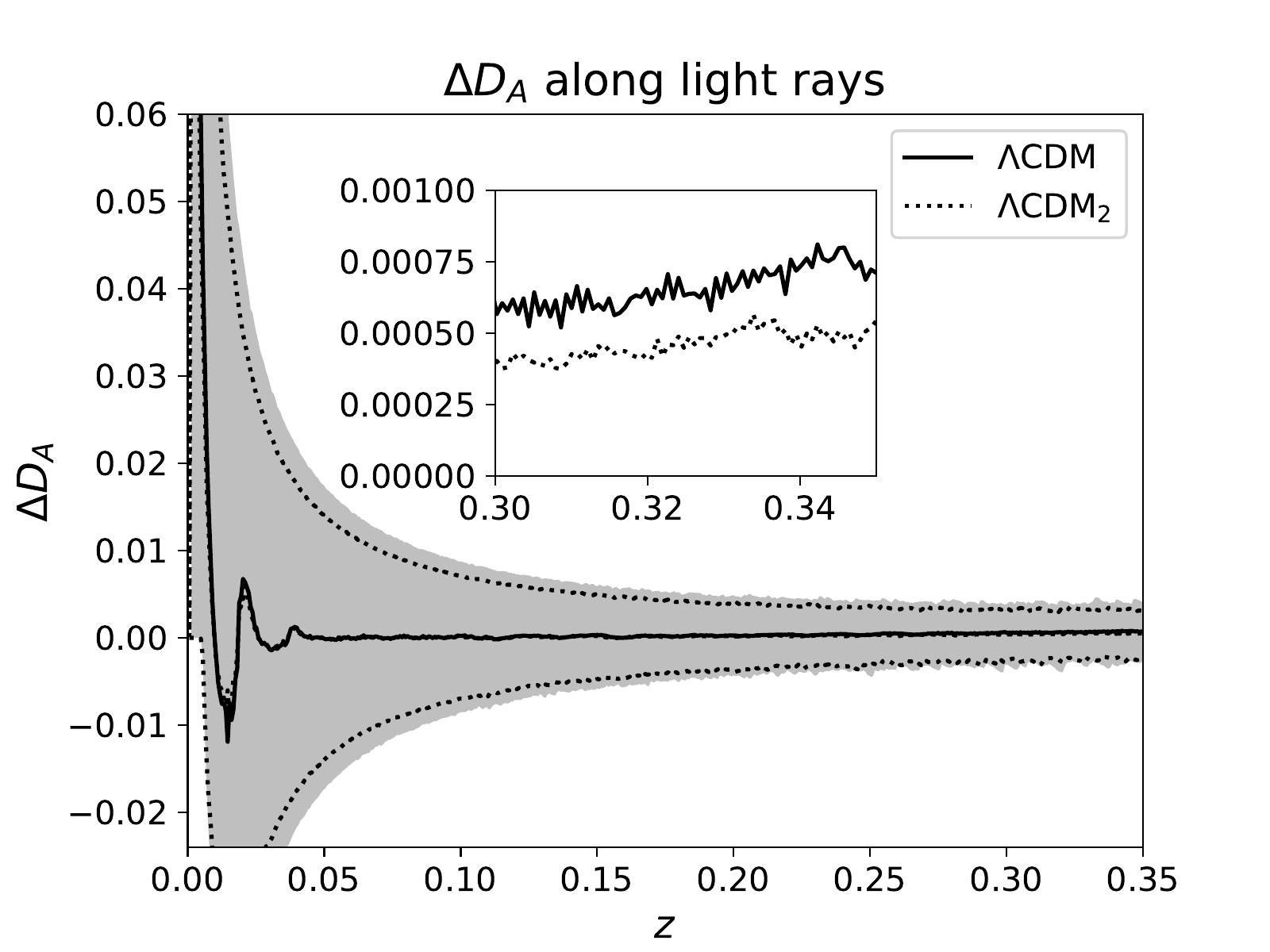}
	}
	\caption{Mean and dispersion of accumulated density contrast, redshift fluctuations and local fluctuations in the angular diameter distance along 1000 light rays in Swiss cheese models with two different flat background models. Shaded areas indicate dispersions amongst light rays within the Swiss cheese model with a standard $\Lambda$CDM background. Close-ups of the mean values are included in the interval $z\in[0.3,0.35]$.}
	\label{fig:flat}
\end{figure}
\begin{figure}
\centering
\includegraphics[scale = 0.5]{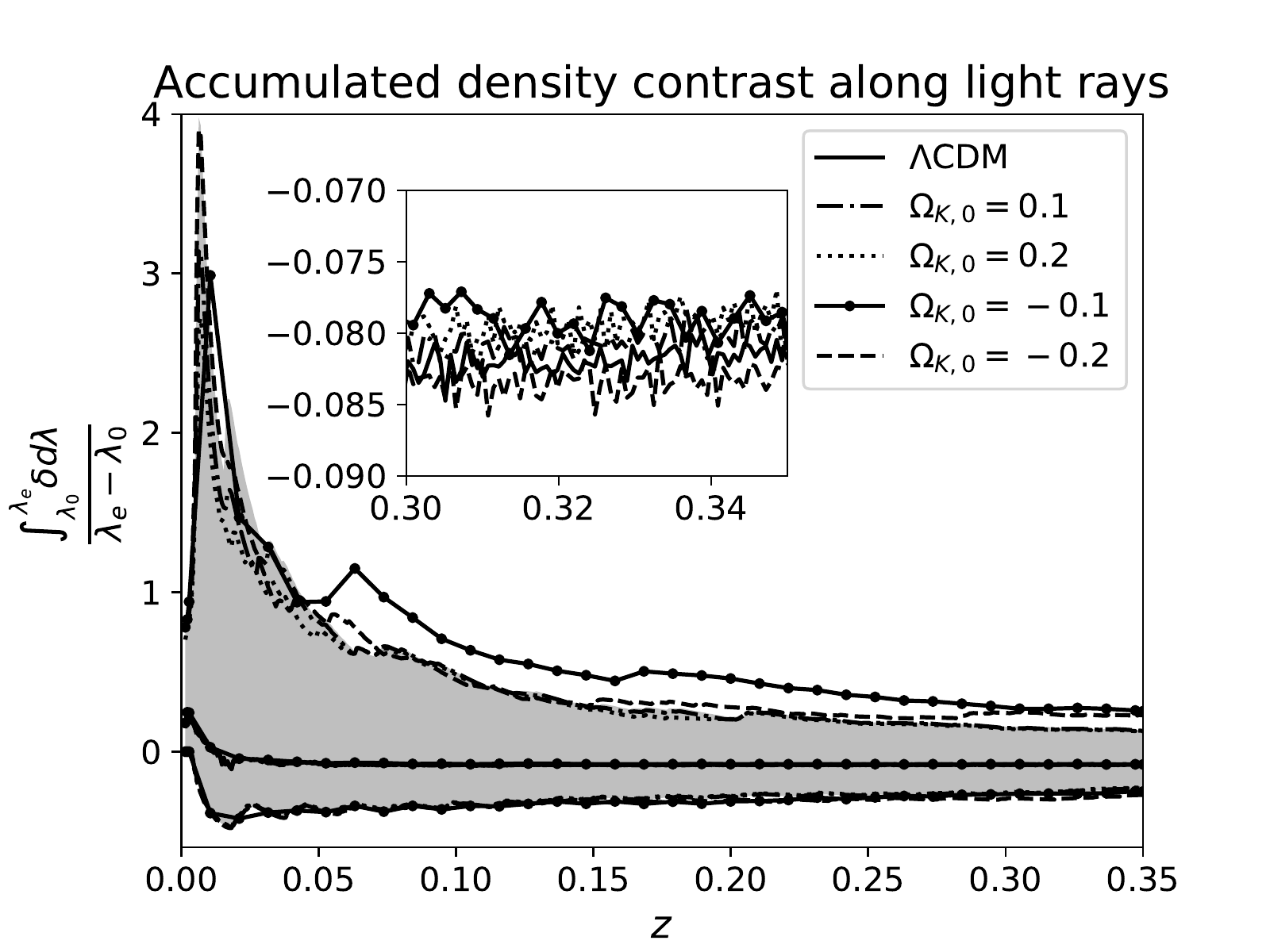}
\caption{Mean and dispersion of accumulated density contrast along 1000 light rays in different Swiss cheese models. The shaded area indicates the dispersion amongst light rays within the Swiss cheese model with a standard $\Lambda$CDM background. Close-ups of the mean values are included in the interval $z\in[0.3,0.35]$.}
	\label{fig:rho}
\end{figure}
\begin{figure}
\centering
\includegraphics[scale = 0.5]{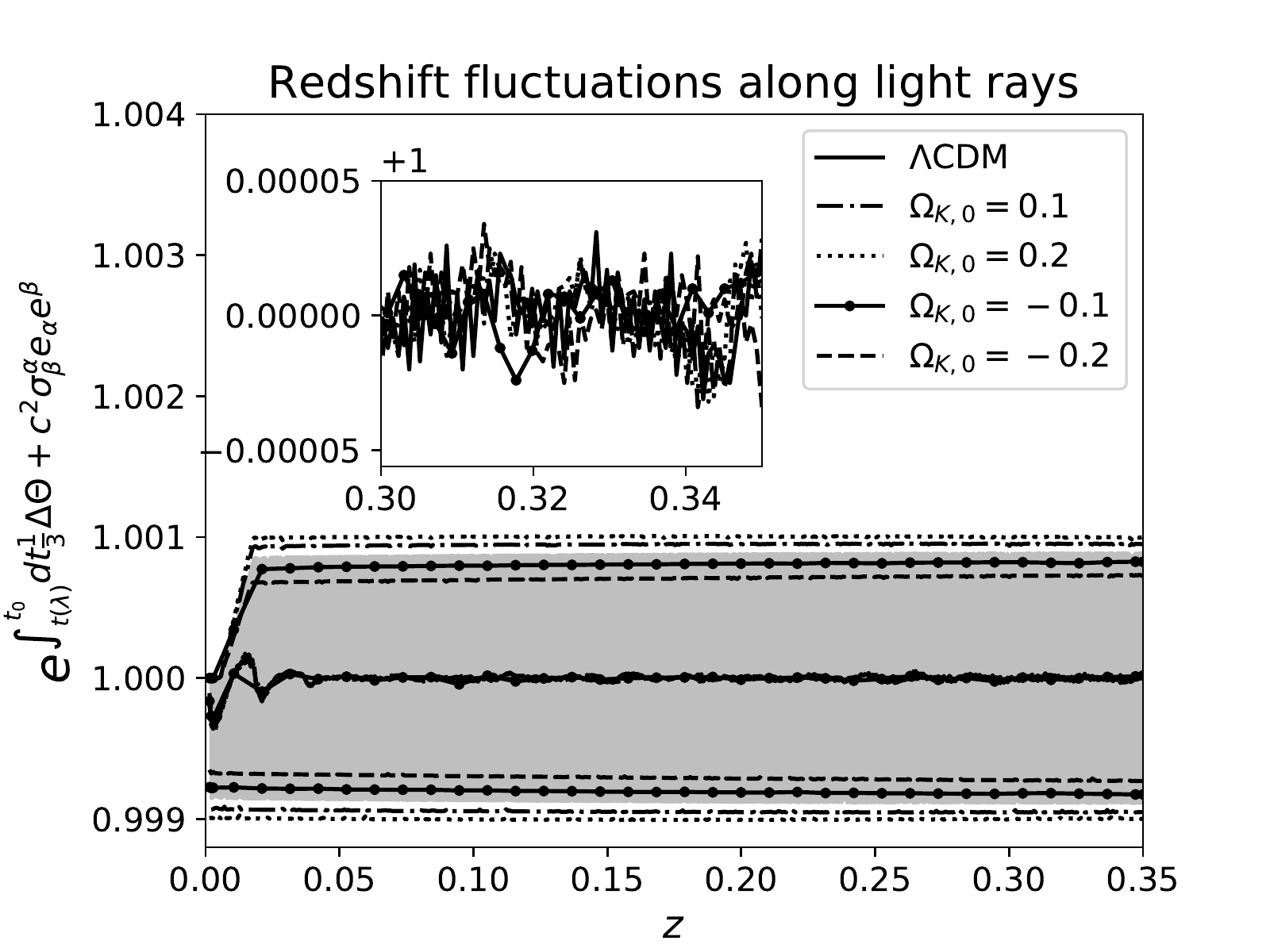}
\caption{Mean and dispersion of fluctuations of the redshift along 1000 light rays in different Swiss cheese models. The shaded area indicates the dispersion for light rays within the Swiss cheese model with a standard $\Lambda$CDM background. Close-ups of the mean values are included in the interval $z\in[0.3,0.35]$.}
\label{fig:z}
\end{figure}
\begin{figure*}
	\centering
	\subfigure[]{
		\includegraphics[scale = 0.5]{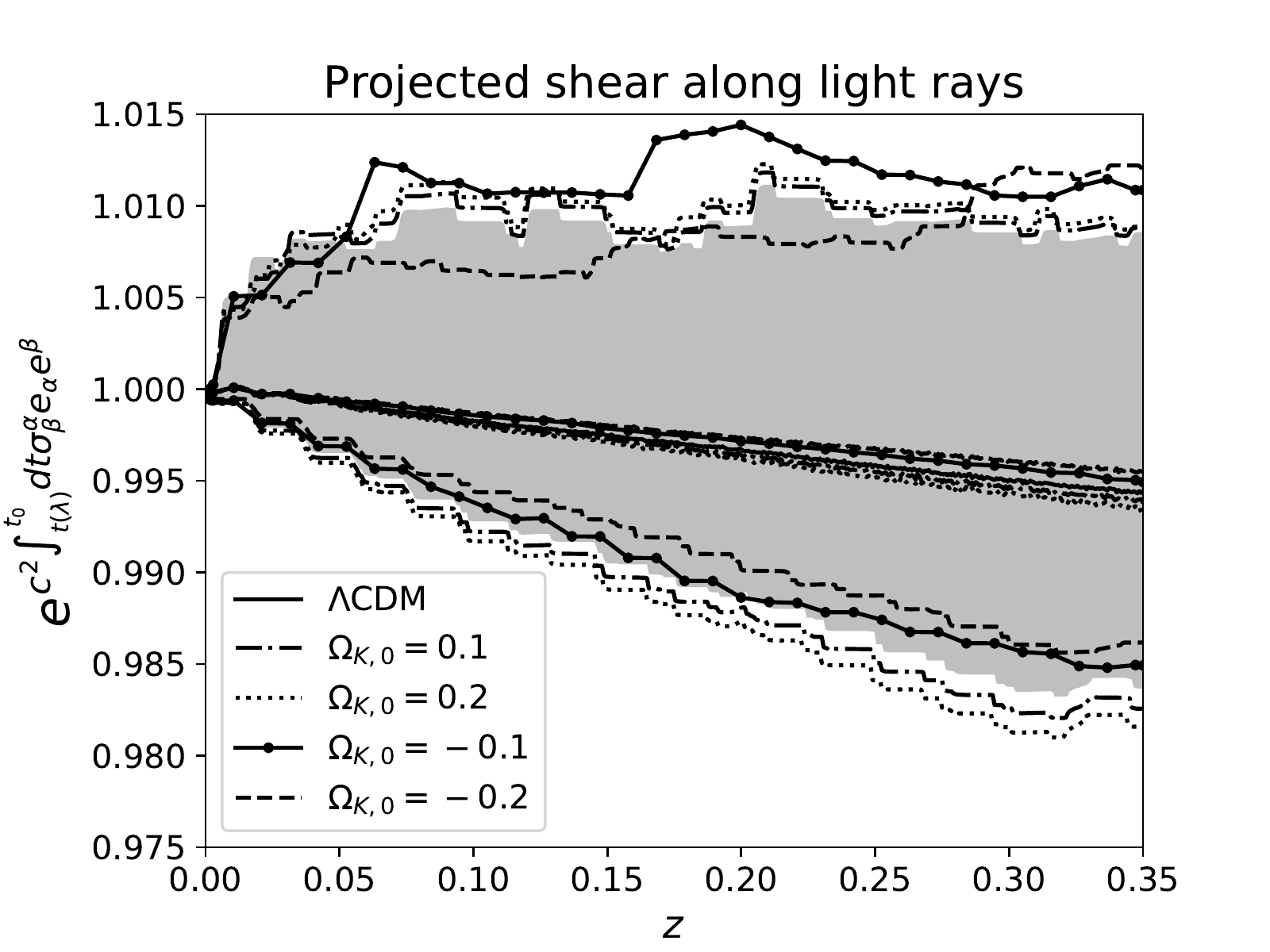}
	}
	\subfigure[]{
		\includegraphics[scale = 0.5]{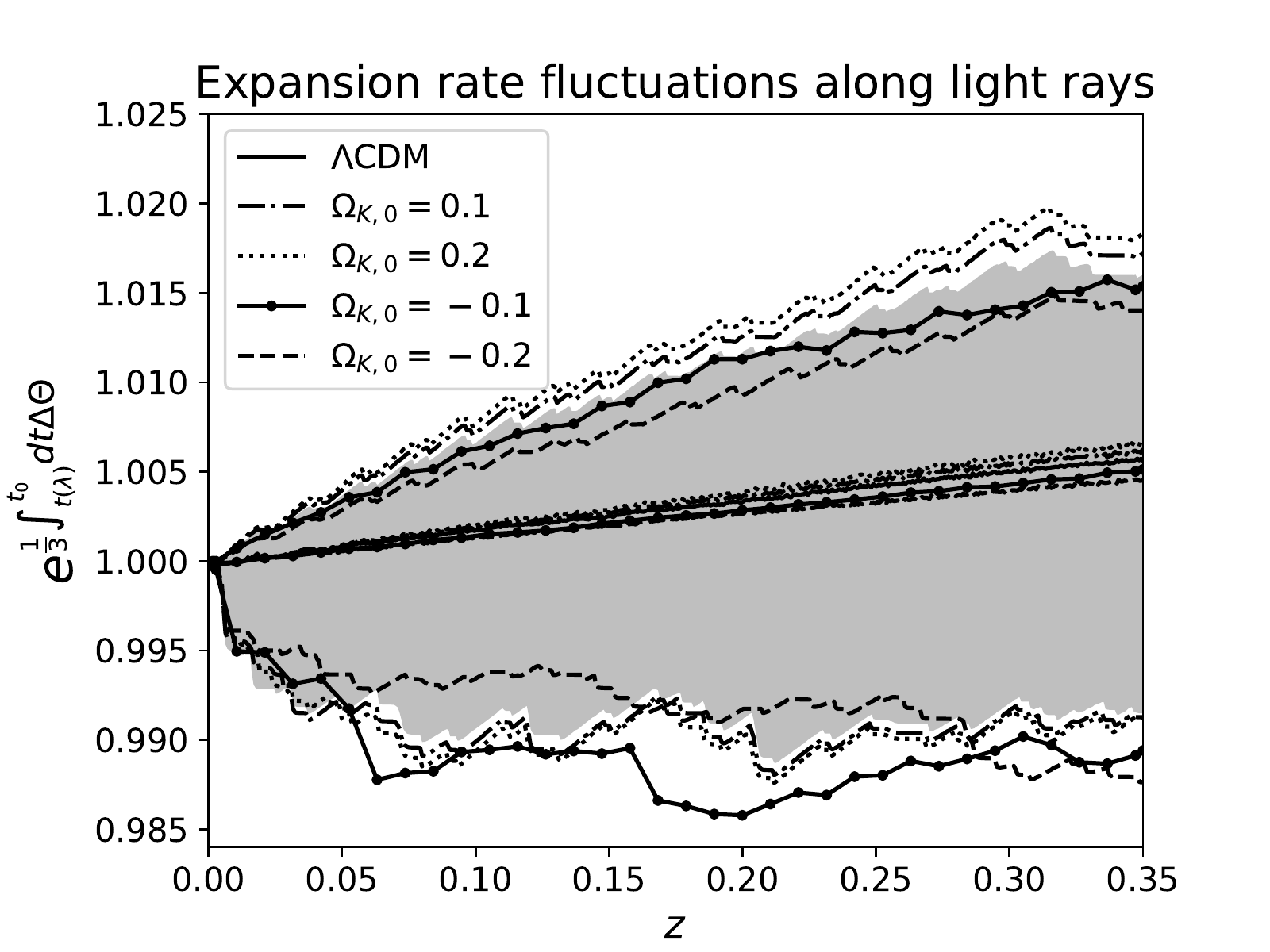}
	}
	\caption{Mean and dispersion of fluctuations in the redshift split into contributions from the projected shear and fluctuations in the expansion rate along 1000 light rays in different Swiss cheese models. The shaded area indicates the dispersion for light rays within the Swiss cheese model with a standard $\Lambda$CDM background.}
	\label{fig:shear}
\end{figure*}
\begin{figure}
\centering
\includegraphics[scale = 0.5]{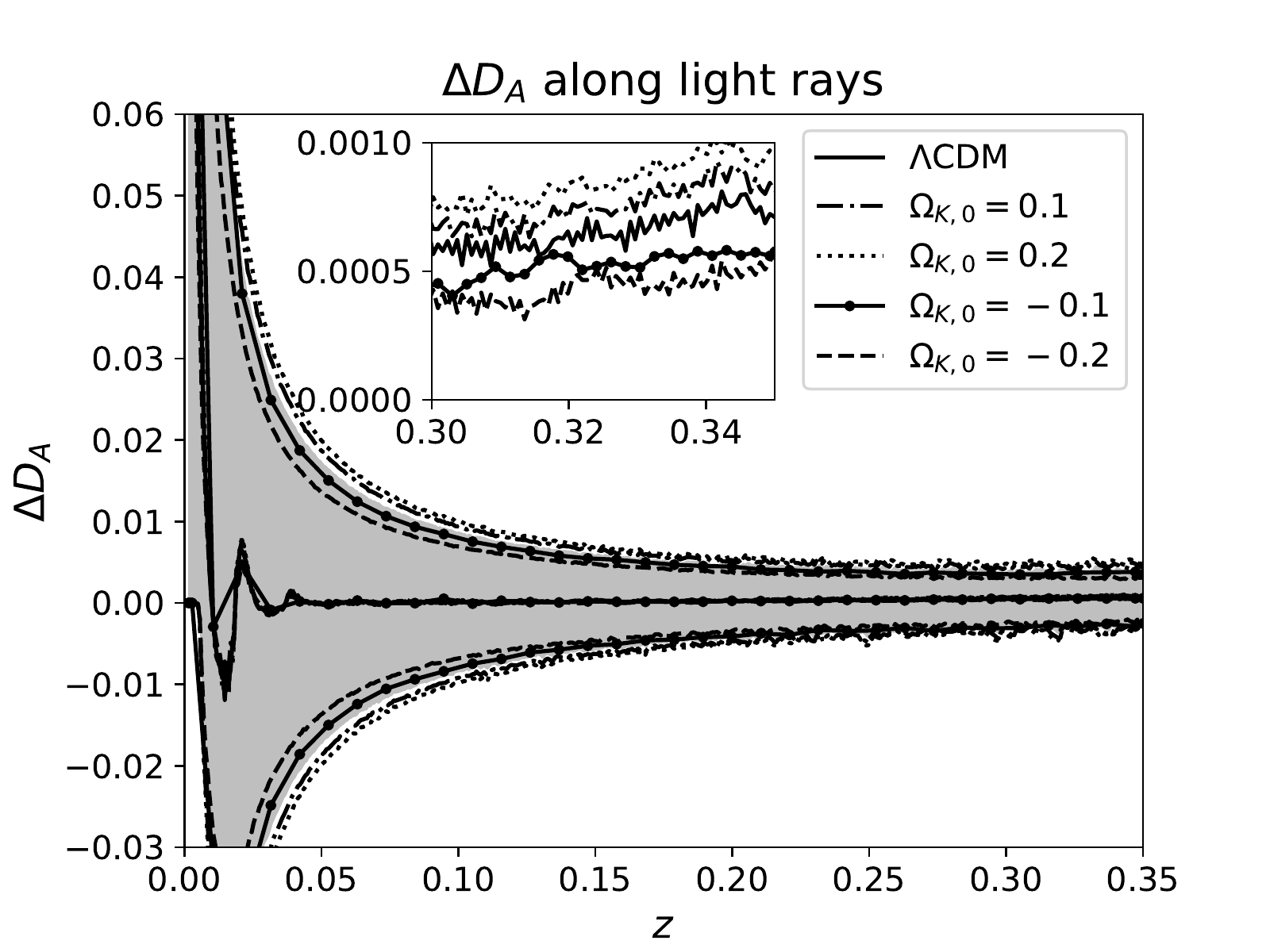}
\caption{Mean and dispersion of $\Delta D_A$ along 1000 light rays in different Swiss cheese models. The shaded area indicates the dispersion for light rays within the Swiss cheese model with a standard $\Lambda$CDM background. Close-ups of the mean values are included in the interval $z\in[0.3,0.35]$.}
\label{fig:DA}
\end{figure}
\begin{figure*}
\centering
\subfigure[]{
\includegraphics[scale = 0.5]{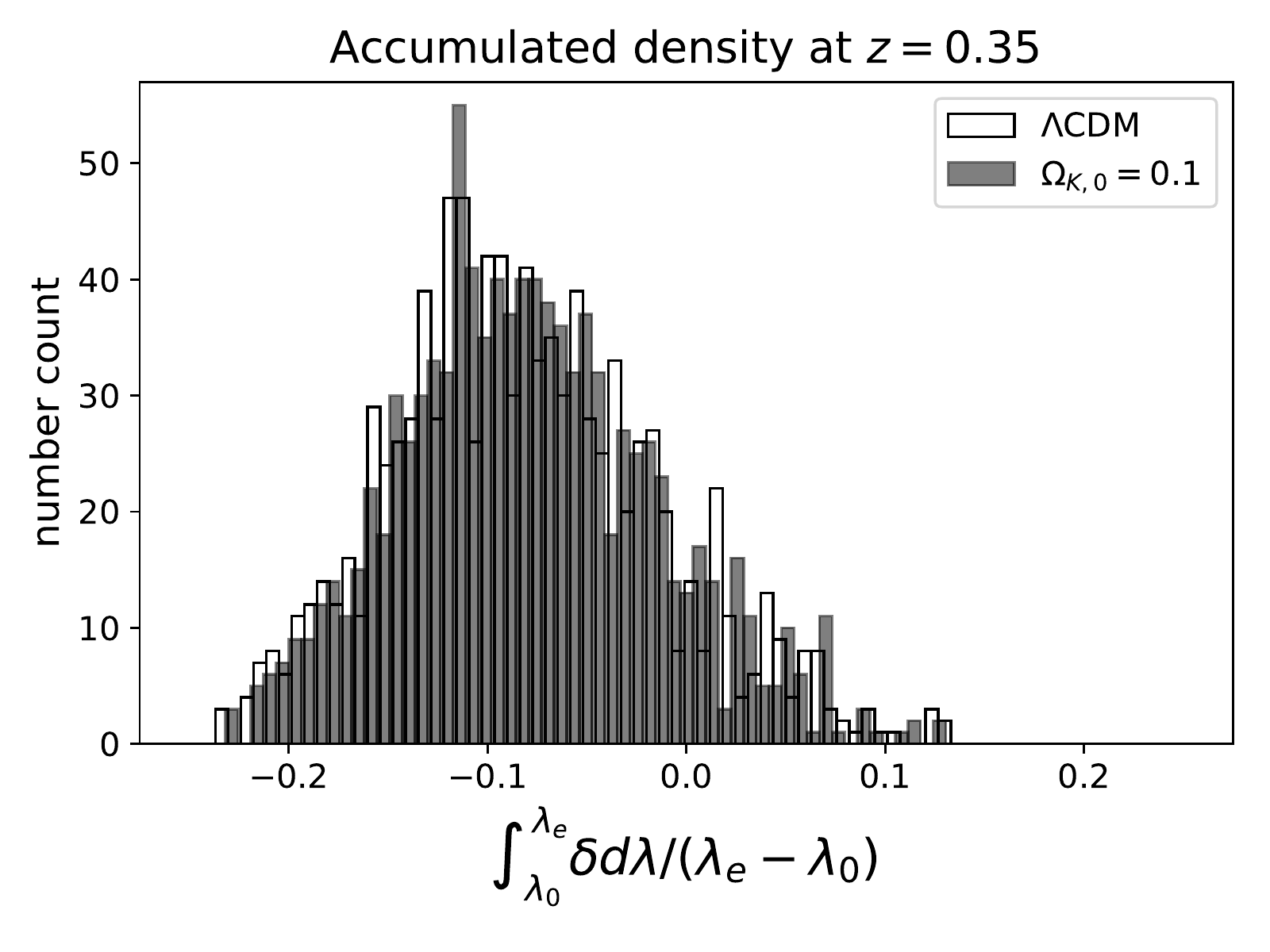}
}
\subfigure[]{
\includegraphics[scale = 0.5]{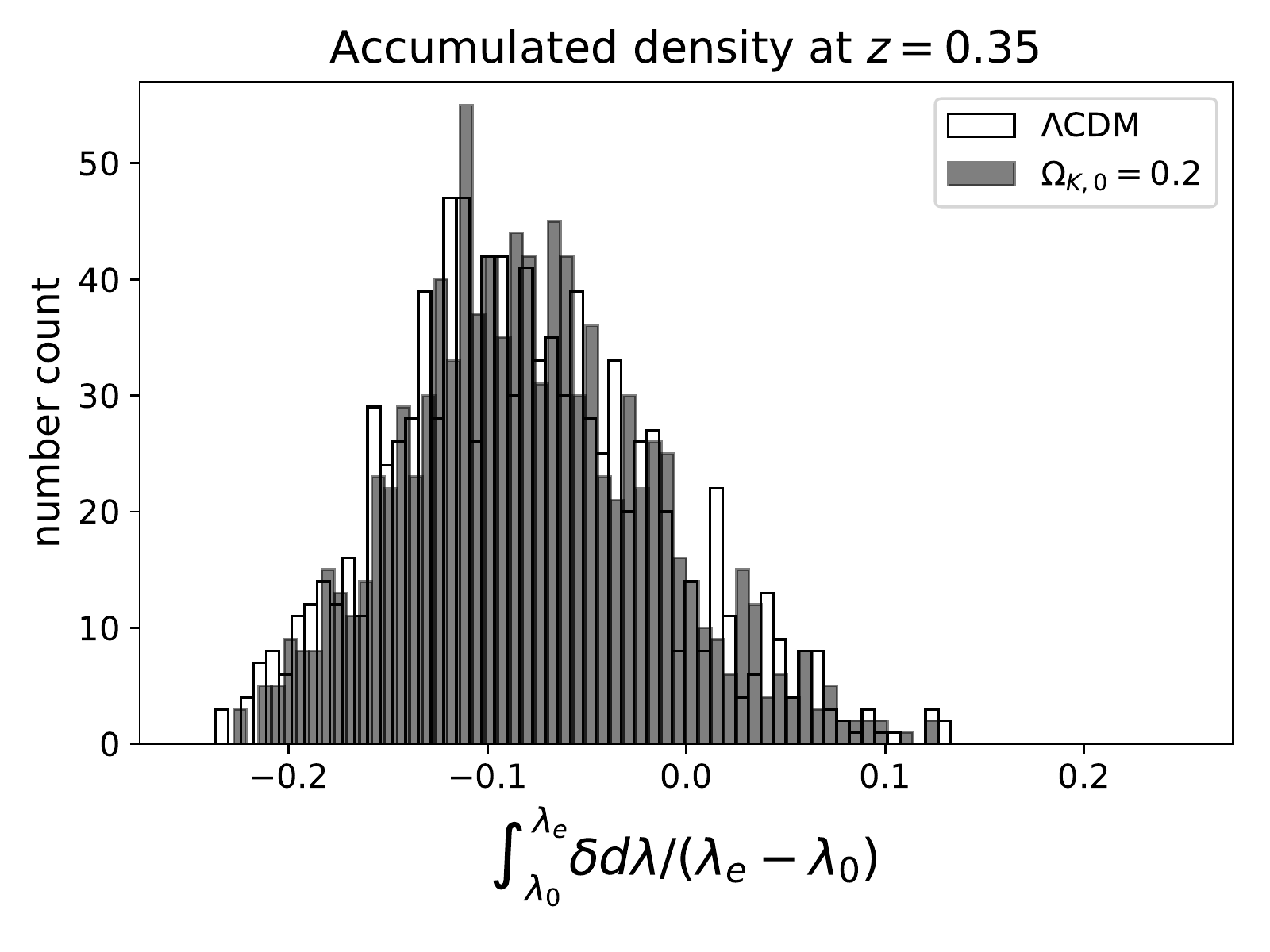}
}\par
\subfigure[]{
\includegraphics[scale = 0.5]{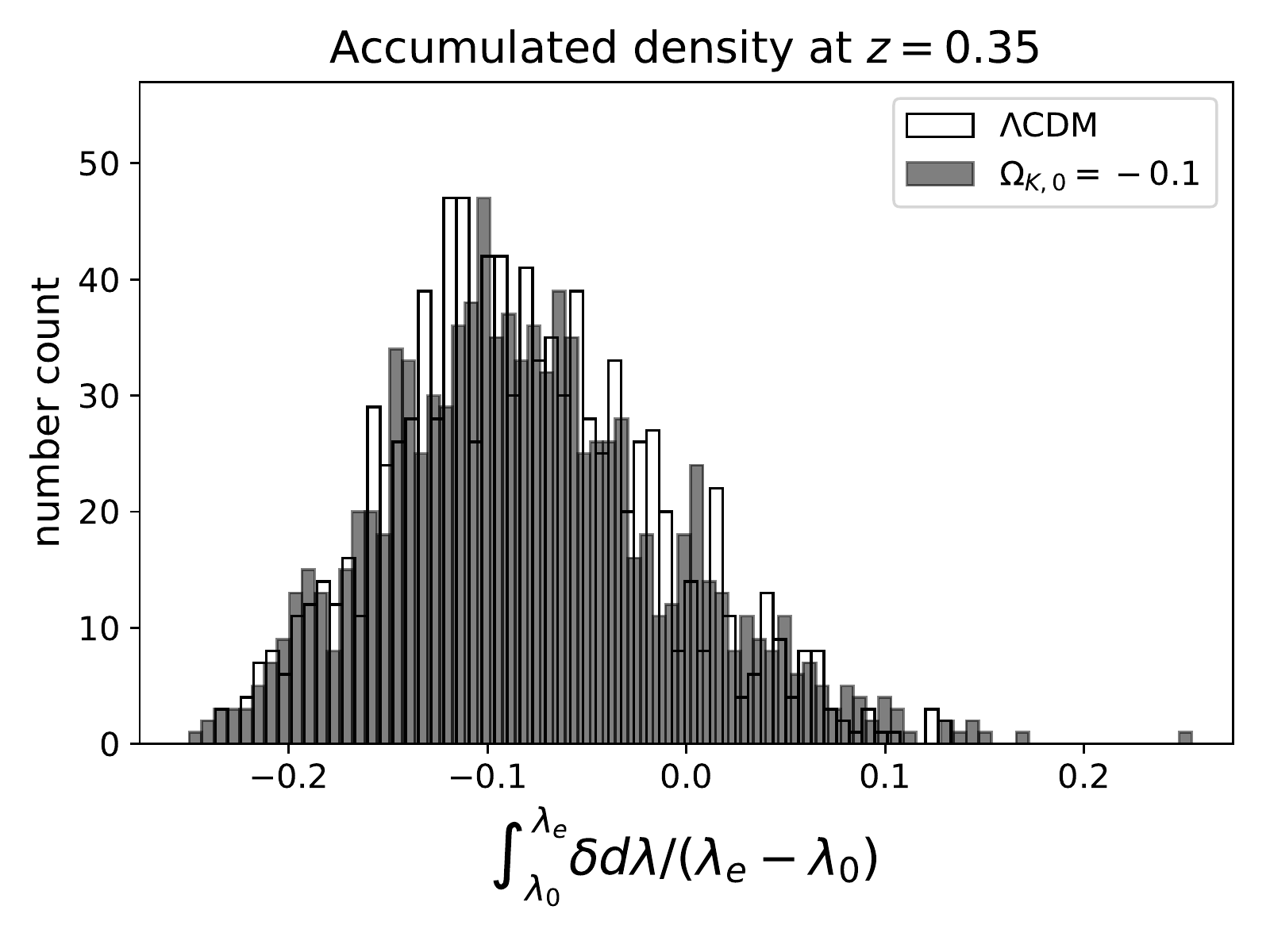}
}
\subfigure[]{
\includegraphics[scale = 0.5]{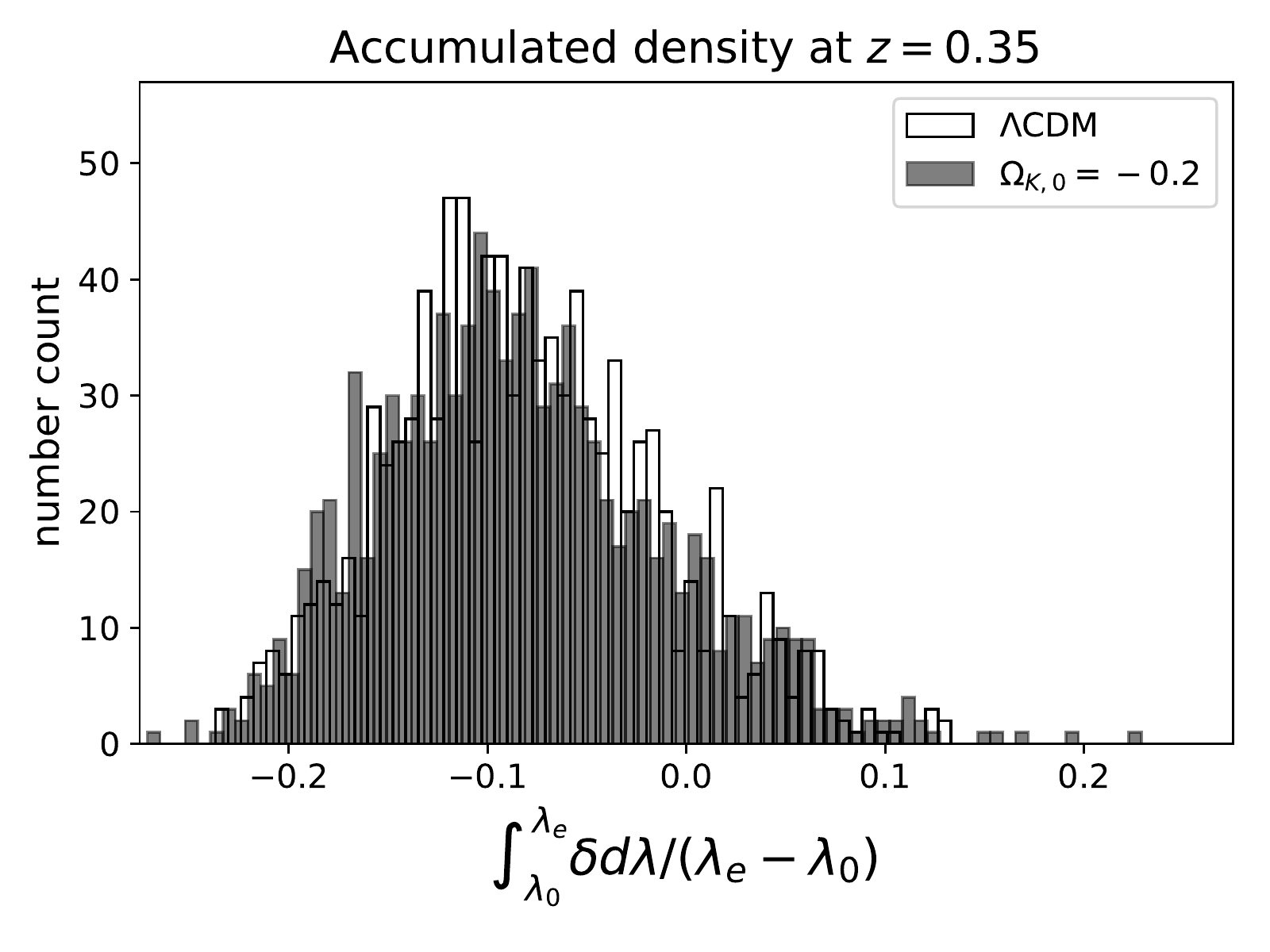}
}
\caption{Histograms showing the accumulated density contrast at $z = 0.35$ for 1000 random light rays in different Swiss cheese models. In each subfigure, results from the model with $\Lambda$CDM as the background is compared with results from a model with curved background. Bin widths are approximately $0.006$. To ease comparison of the individual subfigures, these have all been given the same axis intervals.}
\label{fig:rho_hist}
\end{figure*}
\begin{figure}
\centering
\includegraphics[scale = 0.5]{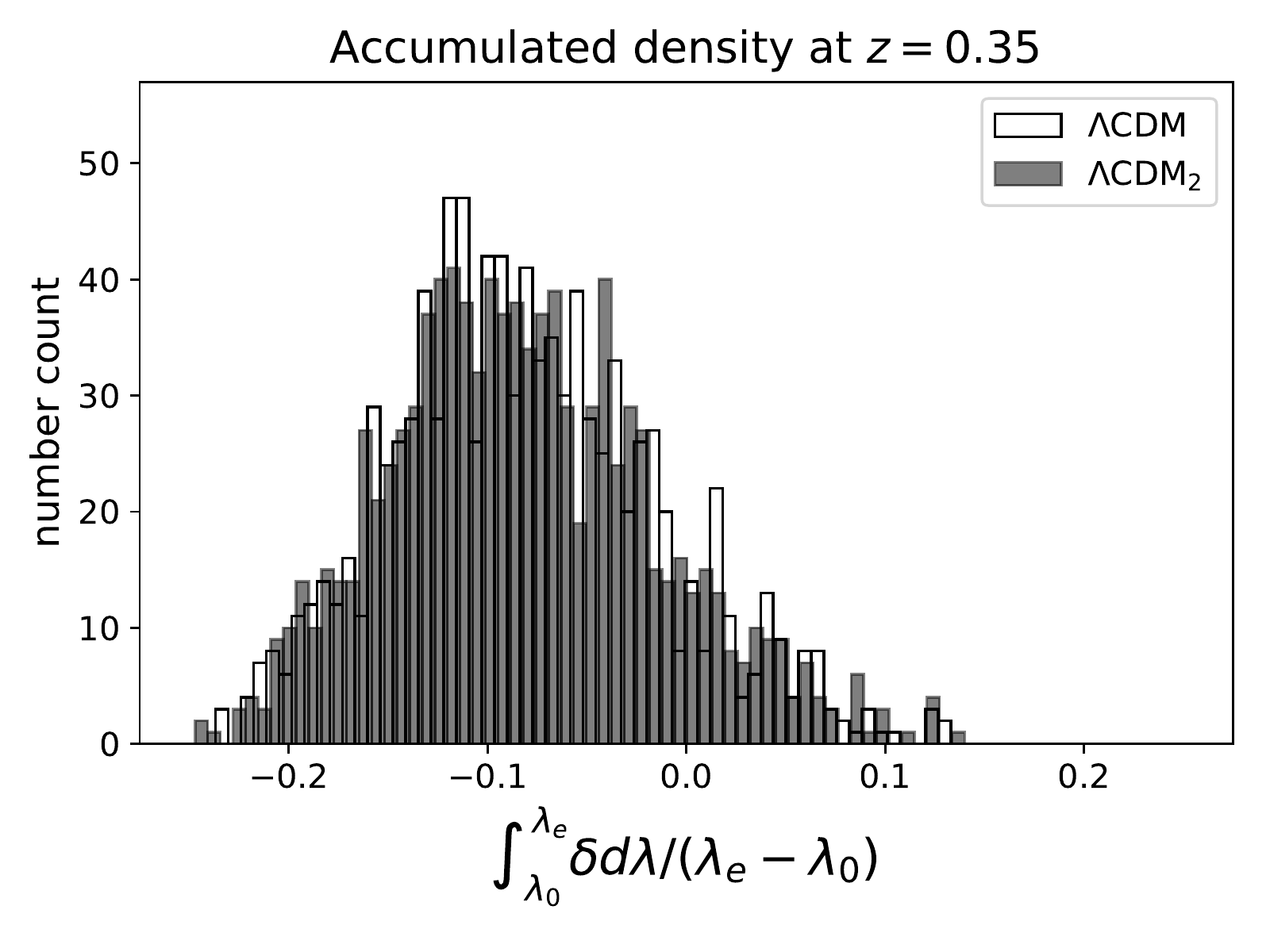}
\caption{Histogram showing the accumulated density contrast at $z = 0.35$ for the two flat models. Bin widths are approximately 0.006.}
\label{fig:rhohist_LCDM}
\end{figure}
\begin{figure*}
\centering
\subfigure[]{
\includegraphics[scale = 0.5]{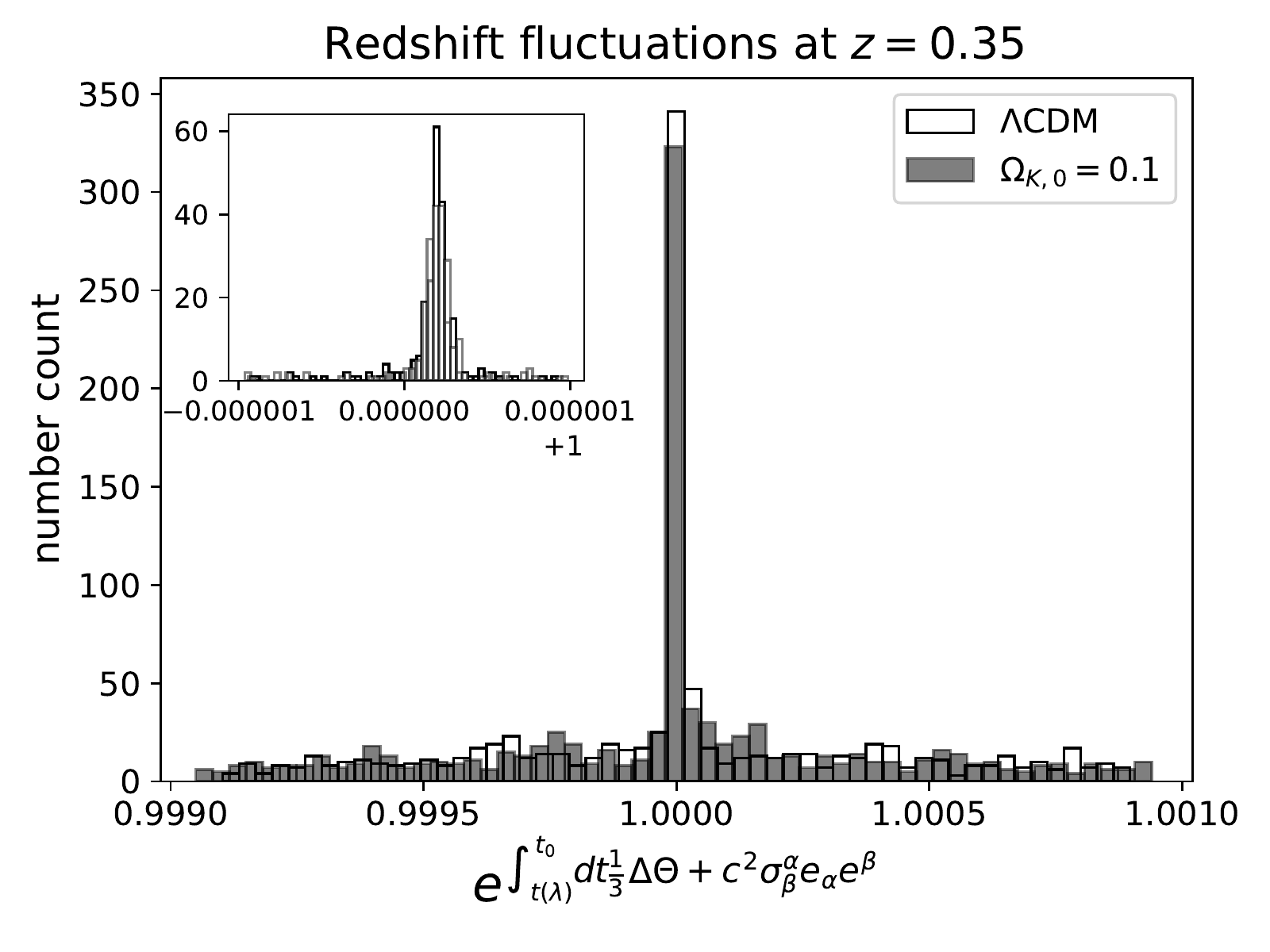}
}
\subfigure[]{
\includegraphics[scale = 0.5]{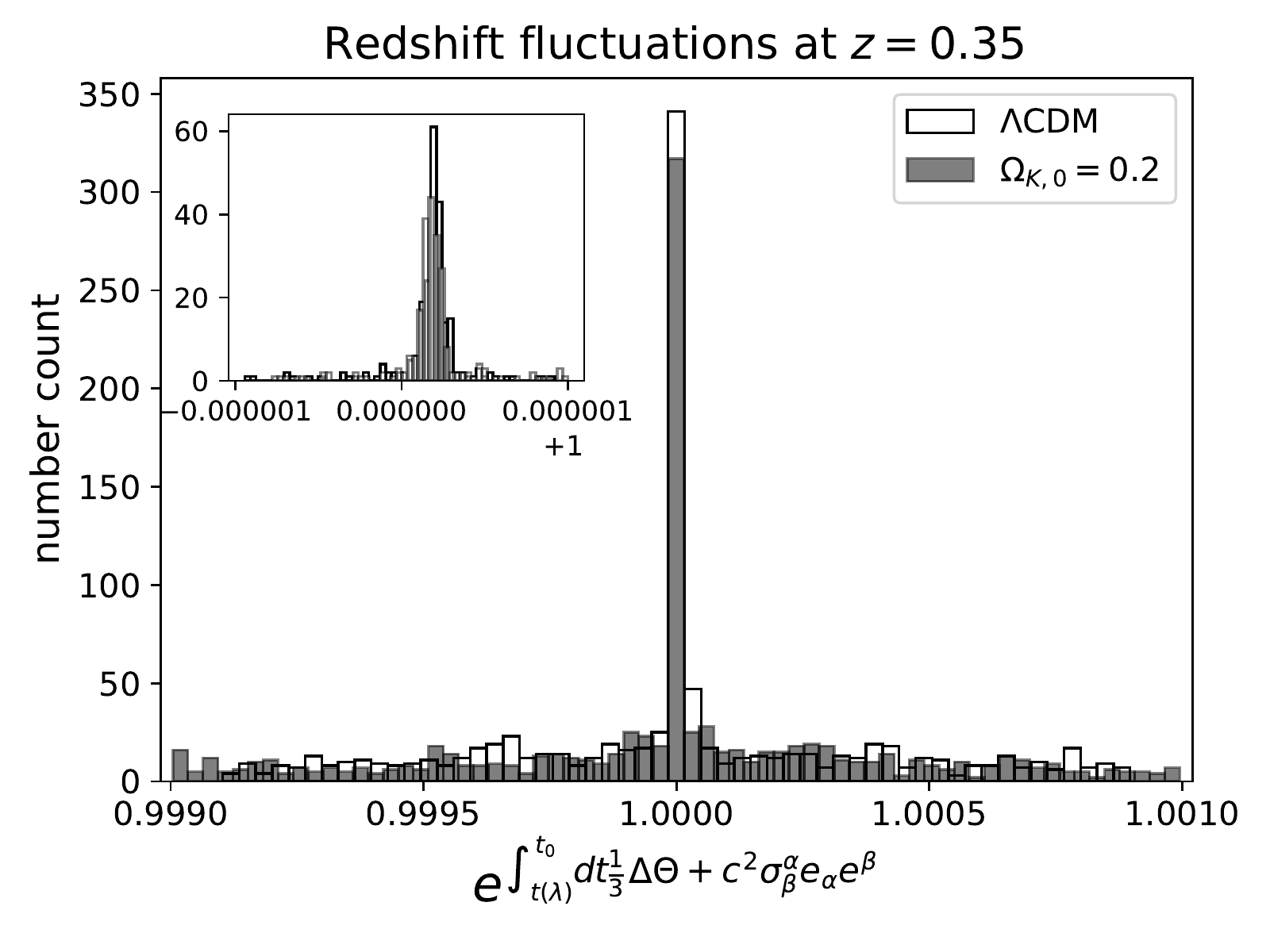}
}\par
\subfigure[]{
\includegraphics[scale = 0.5]{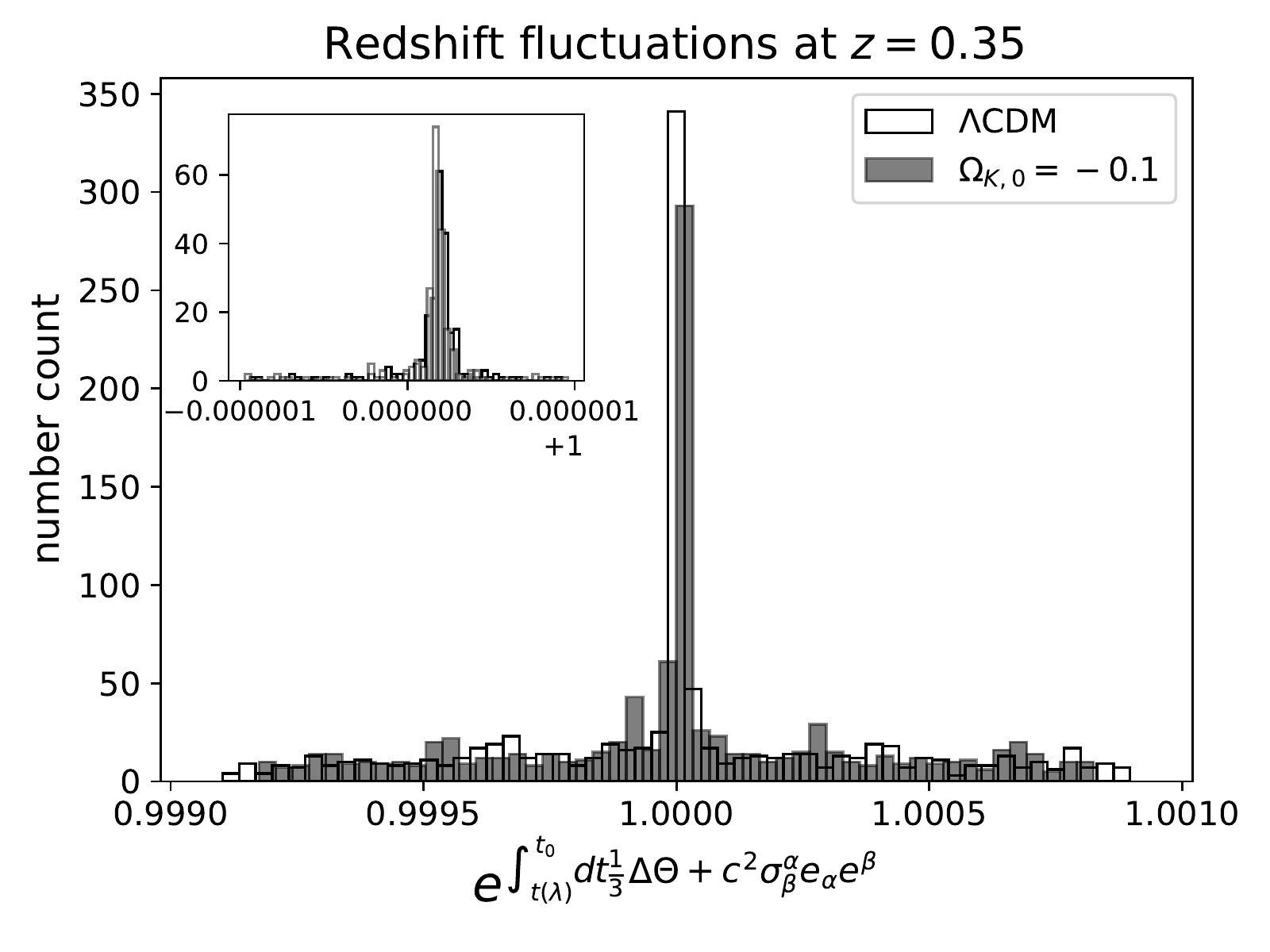}
}
\subfigure[]{
\includegraphics[scale = 0.5]{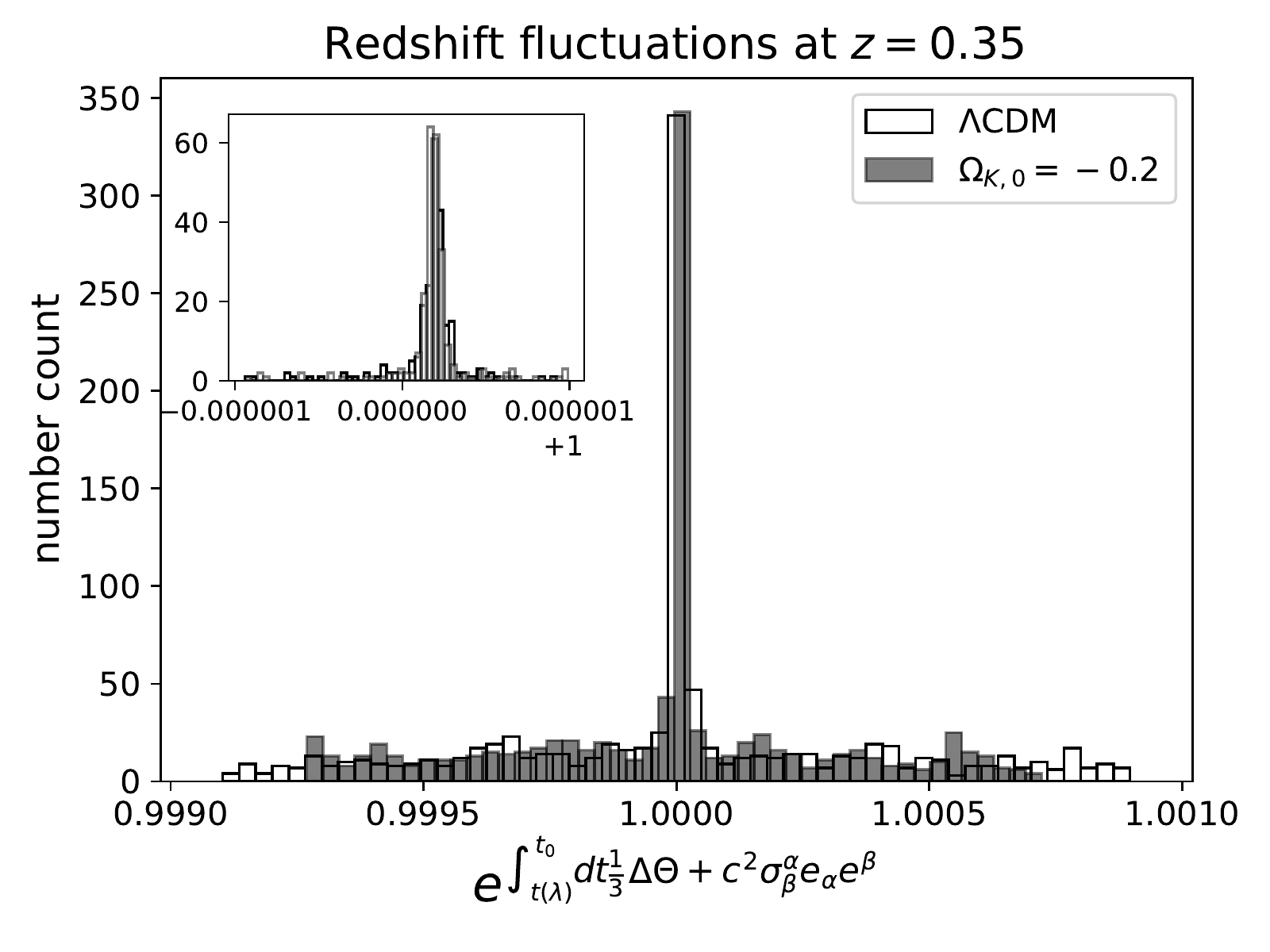}
}\par
\subfigure[]{
\includegraphics[scale = 0.5]{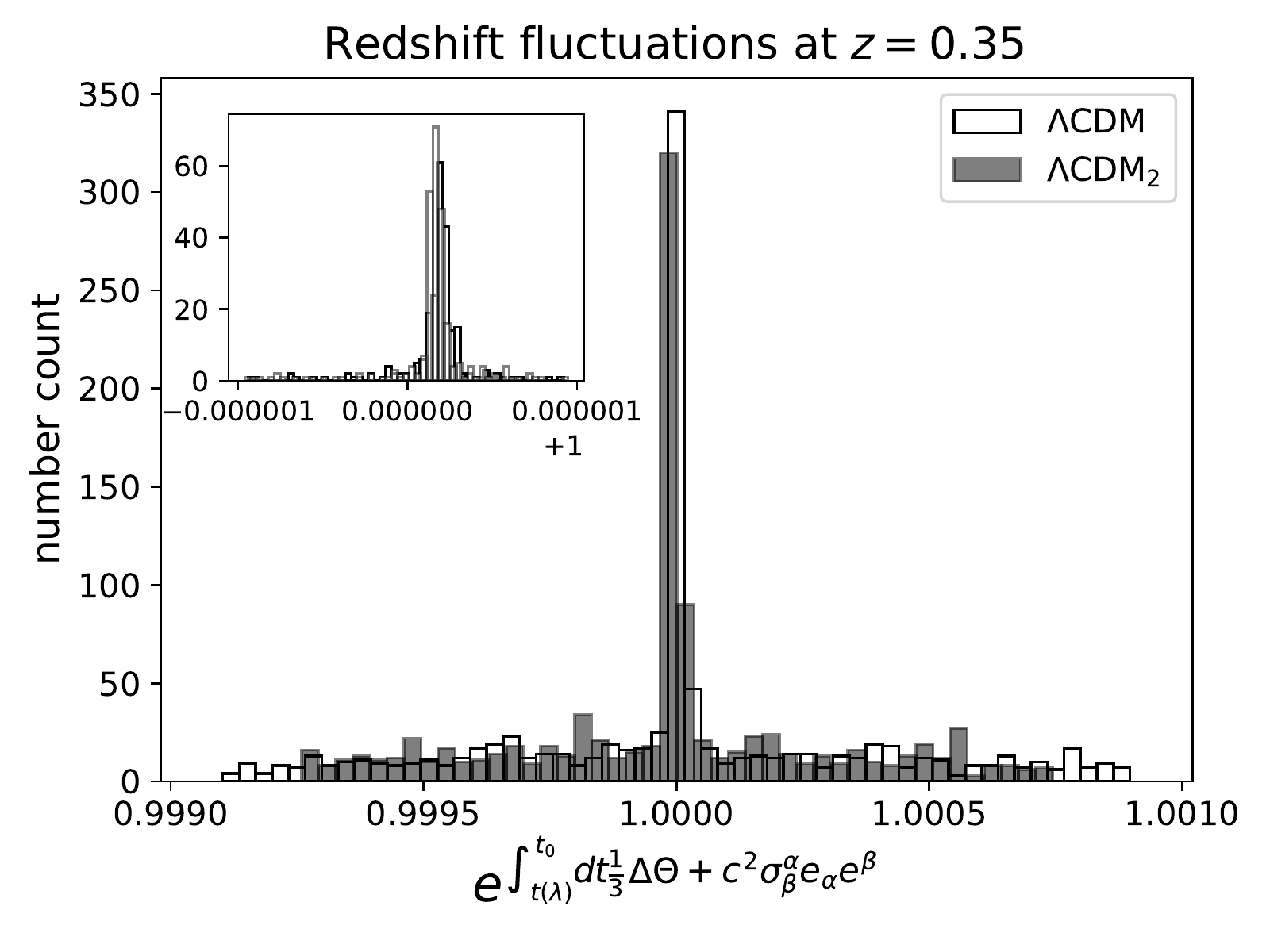}
}
\caption{Histograms showing the redshift fluctuations at $z = 0.35$ for 1000 random light rays in different Swiss cheese models. In each subfigure, results from the model with $\Lambda$CDM as the background is compared with results from a model with curved background. Bin widths are approximately $0.000033$. To ease comparison of the individual subfigures, these have all been given the same axis intervals. Close-ups showing the distributions around the central part of the histograms have been included.}
\label{fig:z_hist}
\end{figure*}
\begin{figure*}
\centering
\subfigure[]{
\includegraphics[scale = 0.5]{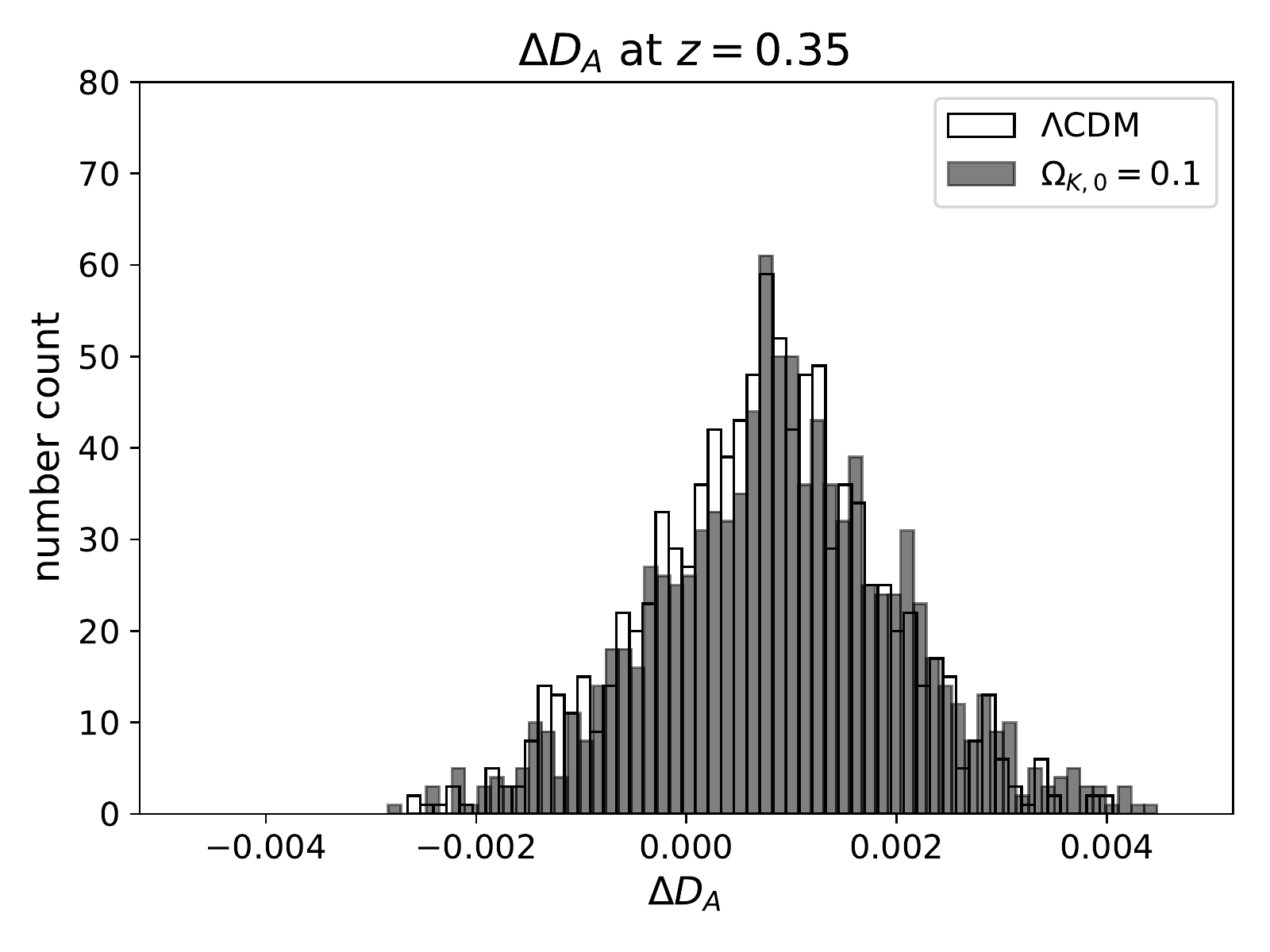}
}
\subfigure[]{
\includegraphics[scale = 0.5]{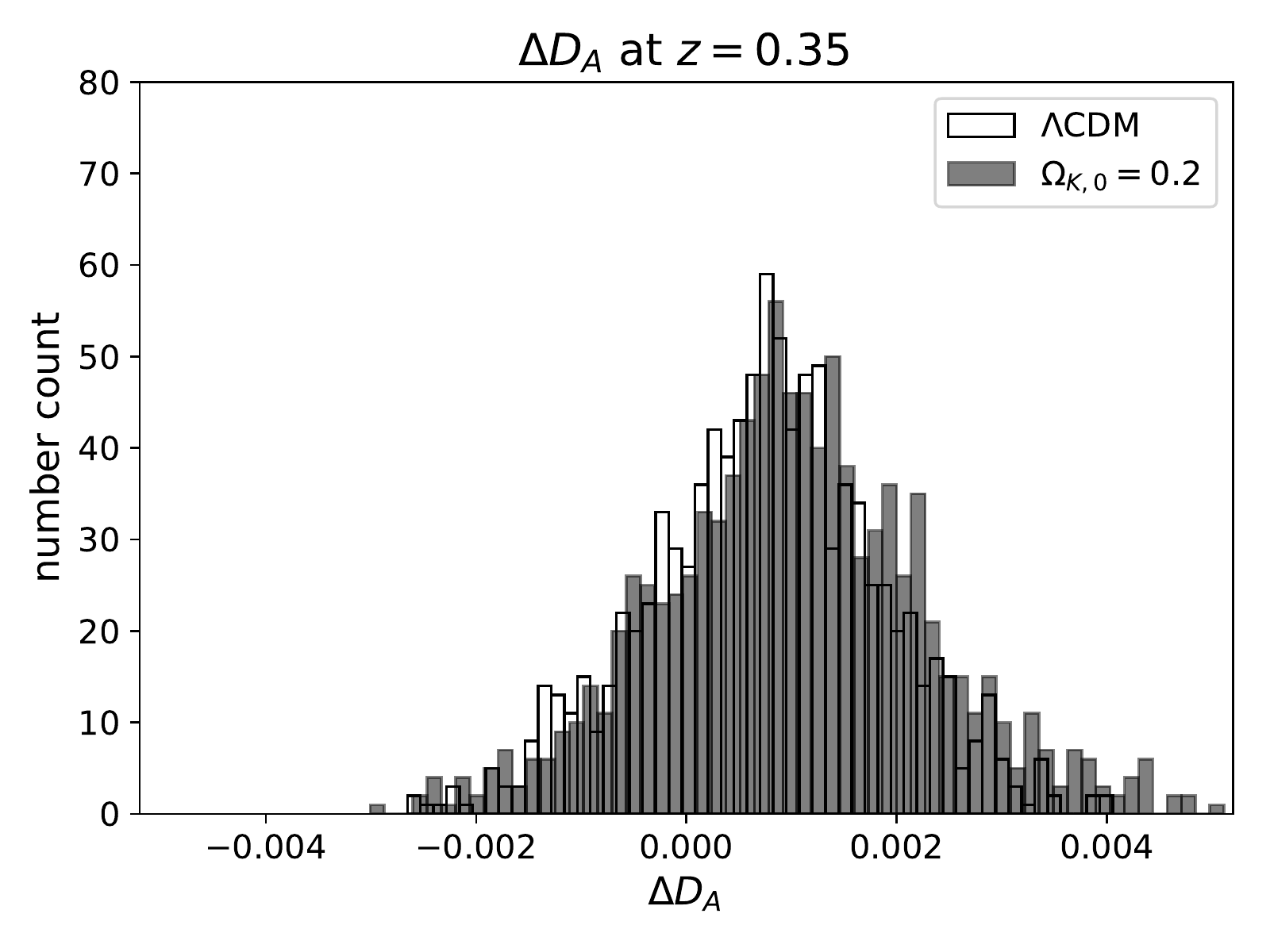}
}\par
\subfigure[]{
\includegraphics[scale = 0.5]{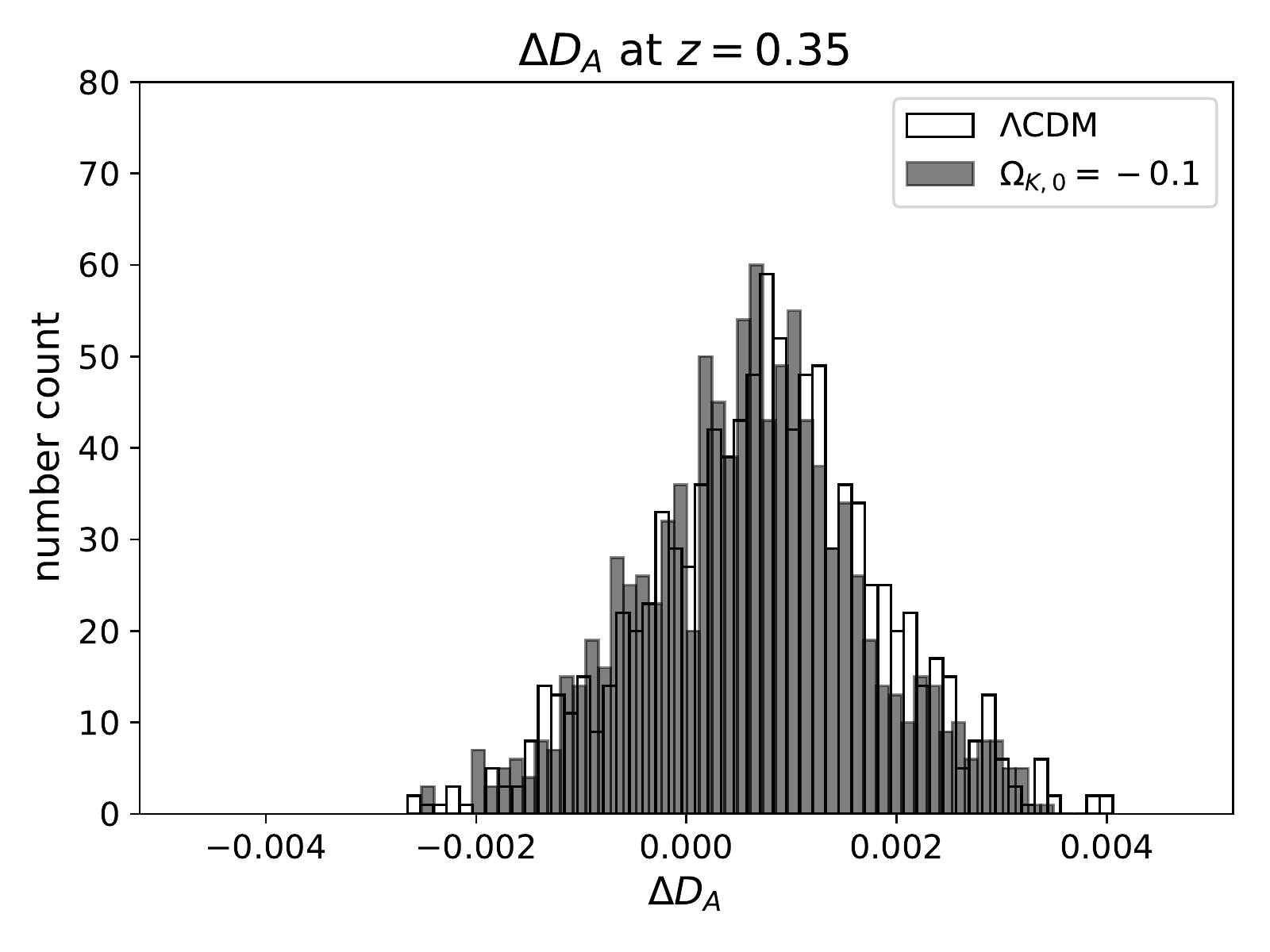}
}
\subfigure[]{
\includegraphics[scale = 0.5]{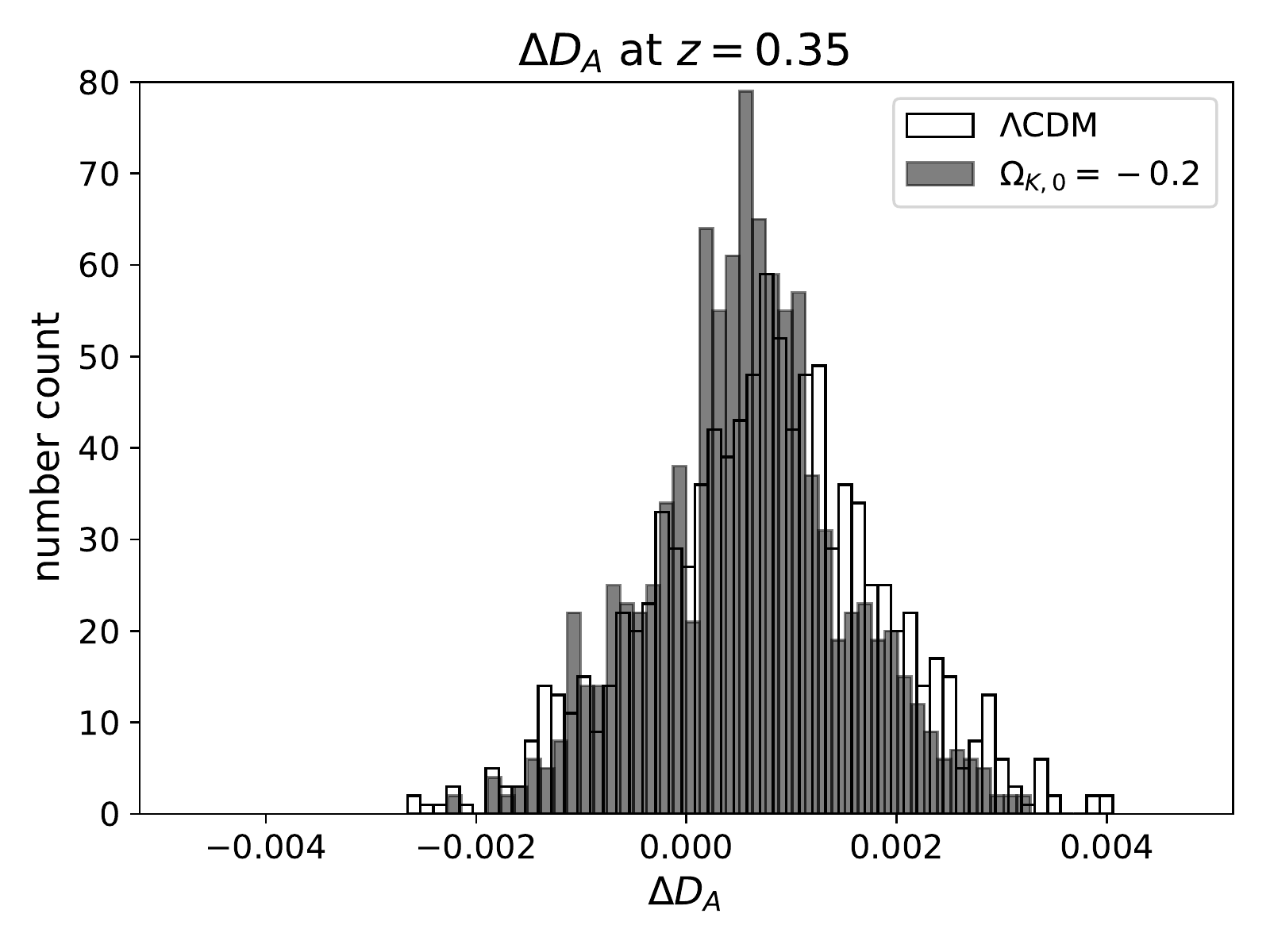}
}\par
\subfigure[]{
\includegraphics[scale = 0.5]{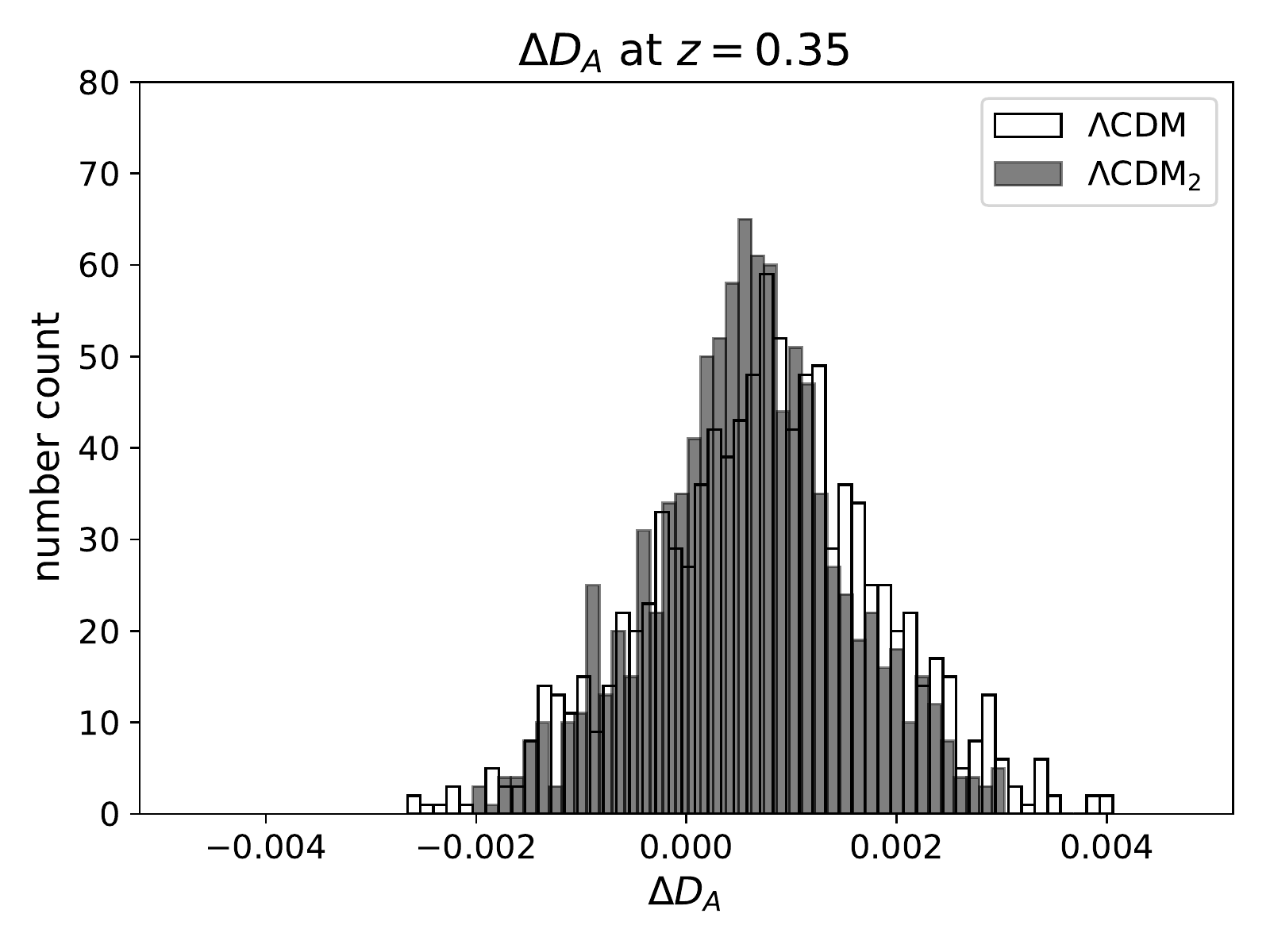}
}
\caption{Histograms showing fluctuations of $D_A$ at $z = 0.35$ for 1000 random light rays in different Swiss cheese models. In each subfigure, results from the model with $\Lambda$CDM as the background is compared with results from a model with curved background. Bin widths are approximately $0.006$. To ease comparison of the individual subfigures, these have all been given the same axis intervals.}
\label{fig:DA_hist}
\end{figure*}
1000 light rays are traced for each model, all with the observer placed at $r = r_b+1$Mpc with random angle between the observer's line of sight and the radial direction towards the LTB structure. Along each light ray, the integrals in equation \eqref{eq:z} are sampled together with the accumulated density contrast, $\frac{\int_{\lambda_0}^{\lambda_e}\delta d\lambda}{\lambda_e-\lambda_0}$, and $\Delta D_A:=\frac{D_A-D_{A,bg}}{D_{A,bg}}$, where $D_{A,bg}$ is the background angular diameter distance. Each light ray is traced until $z = 0.35$ is reached. At this redshift, the light rays have traveled $\gtrsim 1$Gpc and at least for a flat background, the mean values of the sampled quantities should be fairly close to the background values.
\newline\newline
Figure \ref{fig:flat} shows the accumulated density contrast, redshift fluctuations and the local fluctuations in the angular diameter distance for the two flat models. As seen, the mean values are very similar despite the significant difference between the background Hubble parameters and their density contrasts. The dispersions about the mean values are also quite similar, with the dispersion intervals along light rays in the model $\Lambda$CDM$_2$ being slightly smaller for $\Delta D_A$ and $e^{\int_{t(\lambda)}^{t_0}dt\left(\frac{1}{3}\Delta\Theta + c^2\sigma_{\alpha}^{\beta}e^{\alpha}e_{\beta}\right) }$. This is as expected due to the smaller expansion rate of $\Lambda$CDM$_2$ compared to $\Lambda$CDM; a smaller expansion rate makes it easier for local fluctuations in $\Theta$ to cancel along a given light ray.
\newline\indent
It may also be noted that the accumulated density contrasts both converge to approximately $-0.083$ rather than exactly zero. This specific value may be subject to sample variance since only 1000 light rays were considered. To estimate the relevance of sample variance, four other realizations of 1000 light rays in the model with the $\Lambda$CDM background have been studied. The results are shown in appendix \ref{app:variance}. As can be seen in that appendix, it is e.g. found that all five realizations of the 1000 light rays lead to a mean value very close to $-0.08$, deviating from this value at the order of $1-10\%$. The results in the appendix therefore do not indicate that the mean values found here are affected significantly by sample variance. However, it is in principle possible that rare light rays traversing mainly through overdense regions would change the mean value significantly but that they are so rare that they have not been adequately sampled through the 5000 light rays. It is important to note though that effects from such rare light rays are not particularly important here since the point here is not to determine the actual values of the means but rather to compare these values between the different models. As explained in section \ref{subsec:cheese}, the sample variance should be the same for each model studied here.
\newline\newline
The size of the mean value of $\Delta D_A$ seen in figure \ref{fig:flat} is similar to what has been found earlier in studies based on much larger numbers of light rays. For instance, in \cite{cmb_digselv} where nearly $10^5$ light rays were averaged over for each model, the mean value of $\Delta D_A$ at the surface of last scattering was found to be of the order $10^{-4}$ for the models with $p_1=6=p_2$ as here, while the mean value was only of order $10^{-6}$ for models with smaller values of $p_1$ and $p_2$. For the two models studied in \cite{syksyCMB}, mean values of $\Delta D_A$ for approximately 12000 light rays per model at the surface of last scattering were of order $10^{-4}-10^{-5}$. In \cite{Ishak}, 1000 light rays were considered in a Swiss cheese model with quasi-spherical Szekeres structures. The mean fluctuation in $\Delta D_A$ at $z = 1$ was found to be of the order or $10^{-4}$.
\newline\indent
Note again that sample variance may affect the mean through special light rays too rare to have been sampled even in the variance study in appendix \ref{app:variance}. However, the particular value and sign of the mean $\Delta D_A$ is less important than the (dis-)similarities between the values obtained from the two different models. It is here worth mentioning that different signs were found for $\Delta D_A$ in \cite{cmb_digselv,syksyCMB} and that the results in \cite{syksyCMB} were not statistically significant despite averaging over 12000 light rays. The results in \cite{cmb_digselv} were significant to at most 2$\sigma$ despite considering nearly $10^5$ light rays per model. As discussed in appendix \ref{app:variance}, the high packing fractions of the models studied here may lead to relatively high statistical significance in the obtained mean values despite the moderate number of studied light rays per model.
\newline\newline
Figure \ref{fig:rho} shows the mean and dispersion of the accumulated density contrasts along the studied light rays in models with different background curvature. As seen, in all five cases the mean converges to approximately $-0.08$. The maximums and minimums are also very similar for the different models. The dispersion for the model with $\Omega_{K,0} = -0.1$ is, however, prominently larger than the dispersion obtained with the other models. This fits well with the density contrast of that model being largest and does not seem to be due to the background curvature as the dispersion would then be expected to be even more significant for the model with $\Omega_{K,0} = -0.2$.
\newline\indent
As mentioned above, the accumulated density contrast along the lines of sight have been computed as $\frac{\int_{\lambda_0}^{\lambda_e}\delta d\lambda}{\lambda_e-\lambda_0}$. Mainly due to metric measures in spatial averages in curved space, it is not entirely clear how this quantity should be related to neither 1 dimensional spatial averages nor volume averages. Nonetheless, the quantity gives a measure of the accumulated density contrast experienced by the given light ray, i.e. the light path averaged density contrast. If the quantity tends towards the same mean value in each model, then this is an indication that any differences between the light path averaged and the volume averaged density contrasts are not affected by background spatial curvature. The main point with figure \ref{fig:rho} is therefore that, except at very small redshifts, the mean and dispersion of the accumulated density contrasts for the different models are very similar. Deviations at very low redshift are less interesting as they are significantly affected by the exact density contrasts of the individual models.
\newline\newline
The fluctuations in the redshift along the light rays are shown in figure \ref{fig:z}. As seen, the means of the fluctuations are very small. In figure \ref{fig:shear}, these fluctuations are split into contributions from the projected shear and fluctuations in the expansion rate. It is quite striking to see how the two appear to cancel with each other almost exactly - a phenomenon that, as mentioned earlier, has already been demonstrated for other specific LTB models in \cite{Tardis,cmb_digselv}.
\newline\indent
As with the density contrast, the dispersion about the mean behaves somewhat notably for the model with $\Omega_{K,0} = -0.1$ compared to the dispersions of the other models. This could be due to a more significant shearing from the larger density contrast of that model. The larger fluctuations in $\Delta \theta$ must follow from the larger shear if the two contributions are to cancel as appears to be a general phenomenon for many LTB models.
\newline\indent
Figure \ref{fig:DA} shows fluctuations in the angular diameter distance along the light rays in the five models. The picture is again the same as when considering the two flat models: Differences between mean, maximum and minimum of the five models are small and the dispersions become larger as the background expansion rates do. The latter specifically implies that the (small) differences are likely attributable to the models' different expansion rates rather than their background curvatures.
\newline\newline
Before moving on, it is important to stress that a possible presence of e.g. sample variance in the results does not affect the integrity of the study; as already mentioned, the variance should be the same for each model. The main point with the results presented in the figures so far is thus whether or not the mean quantities converge towards the same values for each model - what the actual values are is less interesting. If the means do not converge towards the same values, then this could indicate an effect of background curvature. If, on the other hand, the means of each model {\em do} converge towards the same values, then this indicates that the background curvature does not affect observations on average. This latter possibility seems to be the case when considering the figures discussed so far. 
\newline\newline
Besides looking at the mean values (related to average observations), it is also interesting to look at the distributions around the means. If the distributions have noticeable differences, then this could be an effect of background curvature. Maximums and minimums were shown in the figures discussed so far and showed no indication of effects of background curvature. The actual distributions around the means are also worth considering and are therefore shown in the following. Figure \ref{fig:rho_hist} shows histograms of the distributions of the accumulated density contrast along the light rays at $z = 0.35$. The histogram of each curved model is compared directly with the corresponding histogram of the model with the $\Lambda$CDM background. As seen, the distributions are very similar but become broader and flatter as the background curvature parameter goes from the most positive to the most negative. Especially noticeable is the fact that the two models with negative curvature parameters have small tails on their positive sides. These findings are in good agreement with figure \ref{fig:rho} which also shows that the maximum values are greatest for these two models. Although this result could in principle be a consequence of the background curvature, it seems more likely that the differences seen in the four histograms are due to differences in density contrasts. In particular, the histograms become broader and flatter the larger the present day density contrast of the model is. This is also the case when comparing the two flat models as seen in figure \ref{fig:rhohist_LCDM}.
\newline\newline
Figure \ref{fig:z_hist} shows histograms of redshift fluctuations at $z  = 0.35$. Note specifically that the redshift fluctuations are in comparison to background redshift values (so it makes sense to talk about redshift fluctuations at a specific redshift value). These histograms are very similar, with a large number of light rays in the central histogram bin and broad, flat surrounding distributions. The histograms clearly show that the redshift fluctuations are distributed over slightly larger intervals when going from the most negative to the most positive background curvature parameter. This is in agreement with what should be expected based on the models' background expansion rates which therefore seems the more likely explanation than a curvature effect. Indeed, a comparison of the individual subfigures shows that the model with the $\Lambda$CDM$_2$ background is the most narrow (although the interval is very close to that of model $\Omega_{k,0} = -0.2$). This is also the model with the smallest background expansion rate until present time.
\newline\indent
The histograms in figure \ref{fig:z_hist} are somewhat peculiar looking with a large peak near 1 and the remaining parts of the histograms being flat and broad. The broad, flat parts of the histograms correspond to light rays that are inside LTB structures at $z = 0.35$ while the central parts of the histograms correspond to light rays that are in the background at $z = 0.35$ \footnote{This explanation was suggested by the anonymous referee and later confirmed by the author by looking through the numerical data.}. This prominent difference between the fluctuations inside and outside structures is alluded to in figure 3 of \cite{cmb_digselv} where the redshift fluctuations are traced along individual light rays in Swiss cheese models based on LTB and Szekeres structures. One may also note that the redshift fluctuations are very small when light rays are outside structures, indicating that the spherical symmetry of the LTB models lead to an exceptional cancellation in the integrates Sachs-Wolfe (ISW) \cite{isw} and Rees-Sciama \cite{sciama} effects along light rays. This was also suggested in \cite{syksyCMB} to explain the result that temperature fluctuations in LTB Swiss cheese models were found to be 1-3 orders of magnitude smaller than what would be expected from linear perturbation theory (ISW) and a combination of perturbation theory and N-body simulation data (Rees-Sciama effect). From figure 3 in \cite{cmb_digselv} it can be seen that similar cancellations appear in quasi-spherical Szekeres models so cannot be avoided by using these more complicated models instead of LTB models.
\newline\indent
Figure \ref{fig:z_hist} shows ``close-ups'' of the central parts of the histograms since these are not properly resolved in the main histograms. These close-ups reveal that the redshift fluctuations outside structures are non-vanishing with a distribution that is skewed towards positive fluctuations.
\newline\newline
Lastly, histograms of the distributions of $\Delta D_A$ are shown in figure \ref{fig:DA_hist}. By comparing the six different histograms it is seen that they become (slightly) broader in the sequence $\Lambda$CDM$_2$, $\Omega_{K,0} = -0.2$, $\Omega_{K,0} = -0.1$, $\Lambda$CDM, $\Omega_{K,0} = 0.1$, $\Omega_{K,0} = 0.2$. This is exactly the sequence of increasing expansion rate and the histogram features are thus most likely attributable to this factor. Hence, as with the other histograms, no effect that can convincingly by attributed the background curvature is found.
\newline\newline
Overall, the results presented in this section do not indicate any significant effects of the background spatial curvature on the relation between volume and light path averages that could be significant for the interpretation of observations.
\newline\indent
If very determined to identify a possible effect of background curvature, the most promising, based on the presented result, seems to be the very small tails towards high values of the accumulated density contrast seen for the two models with negative curvature parameters. However, these tails seem to be naturally attributable to statistical flukes of tracing mainly overdensities along some light rays. Such a situation would lead to more prominent tails in the models with larger density contrasts and might not be visible in the models with smaller density contrasts. In relation to this suggested explanation, it may be noted that similar tails were found in two of the studied realizations of 1000 light rays in the model with the $\Lambda$CDM background. This can be seen in histograms presented in appendix \ref{app:variance} where a discussion of this finding is also given.

\section{Summary}\label{sec:Summary}
It was remarked in \cite{Tardis} that it is unclear whether or not 1 dimensional spatial averages converge towards volume averages if space is not Euclidean. In a statistically homogeneous and isotropic spacetime with slowly evolving structures, light paths can be approximated through 1 dimensional spatial averages. Hence, if 1 dimensional averages do in fact not converge to volume averages in curved space, it can lead to important biases in observations if the real Universe has a small, non-vanishing curvature. This possibility was here studied by computing the redshift-distance relation, accumulated density contrast and redshift fluctuations through the integrated expansion rate and shear along 1000 light rays in each of a series of Swiss cheese models with LTB structures and FLRW backgrounds of different spatial curvature.
\newline\indent
Small differences between the models in mean values and the dispersion of the computed quantities were found. These differences were identified as being likely due to small differences in expansion rates and density contrasts rather than to the background curvatures being different. This assessment is supported by comparing to the same computations in two models with different flat backgrounds. Thus, the results presented here indicate that results obtained for models with flat backgrounds regarding e.g. mean and dispersion in $H_0$ studied in relation to the $H_0$-problem are valid even if the Universe has a small non-vanishing curvature (possibly emerging only at late times due to cosmic backreaction).

\section{Acknowledgments}
The author would like to thank Steen Hannestad for facilitating the use of cluster computer resources. Brigitte Henderson is also thanked for her help with the necessary vpn connection. Syksy Rasanen is thanked for conversations and explanations regarding the issue of 1 dimensional versus volume averages in curved space. Lastly, the anonymous referee is thanked for several comments and suggestions which have improved the presentation of the work significantly.
\newline\indent
The numerical results presented in this work were obtained at the Centre for Scientific Computing, Aarhus http://phys.au.dk/forskning/cscaa/.
\newline\newline
The author is supported by the Independent Research Fund Denmark under grant number 7027-00019B.

\FloatBarrier
\appendix
\section{Sample variance}\label{app:variance}
\begin{figure}
\centering
\includegraphics[scale = 0.5]{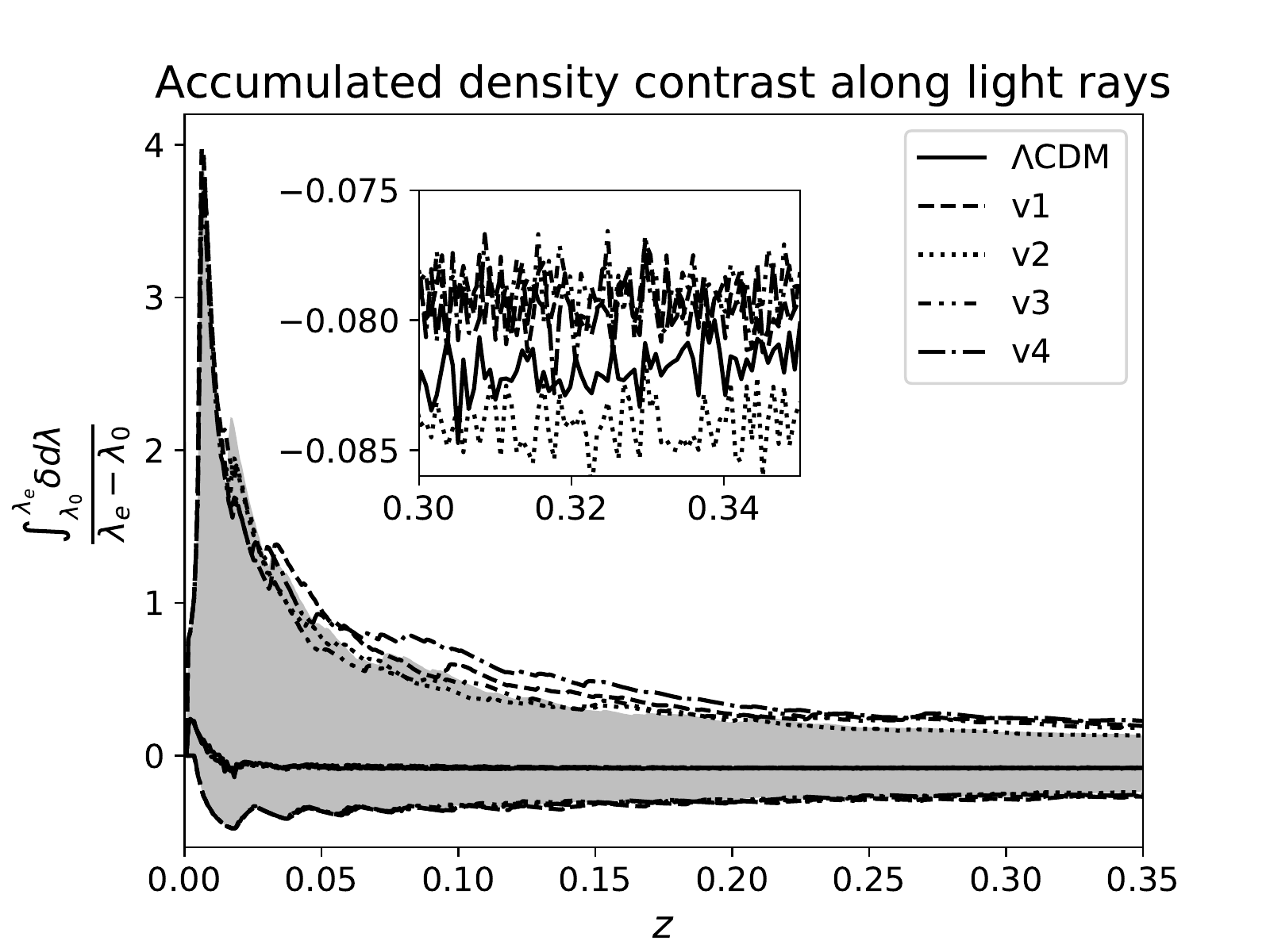}
\caption{Mean and dispersion of the accumulated density contrast along light rays based on five different realizations of 1000 random light rays. The shaded area indicates the dispersion amongst light rays with the same realization as that used in the main text. Close-ups of the mean values are included in the interval $z\in[0.3,0.35]$.}
	\label{fig:rho_var}
\end{figure}

\begin{figure}
\centering
\includegraphics[scale = 0.5]{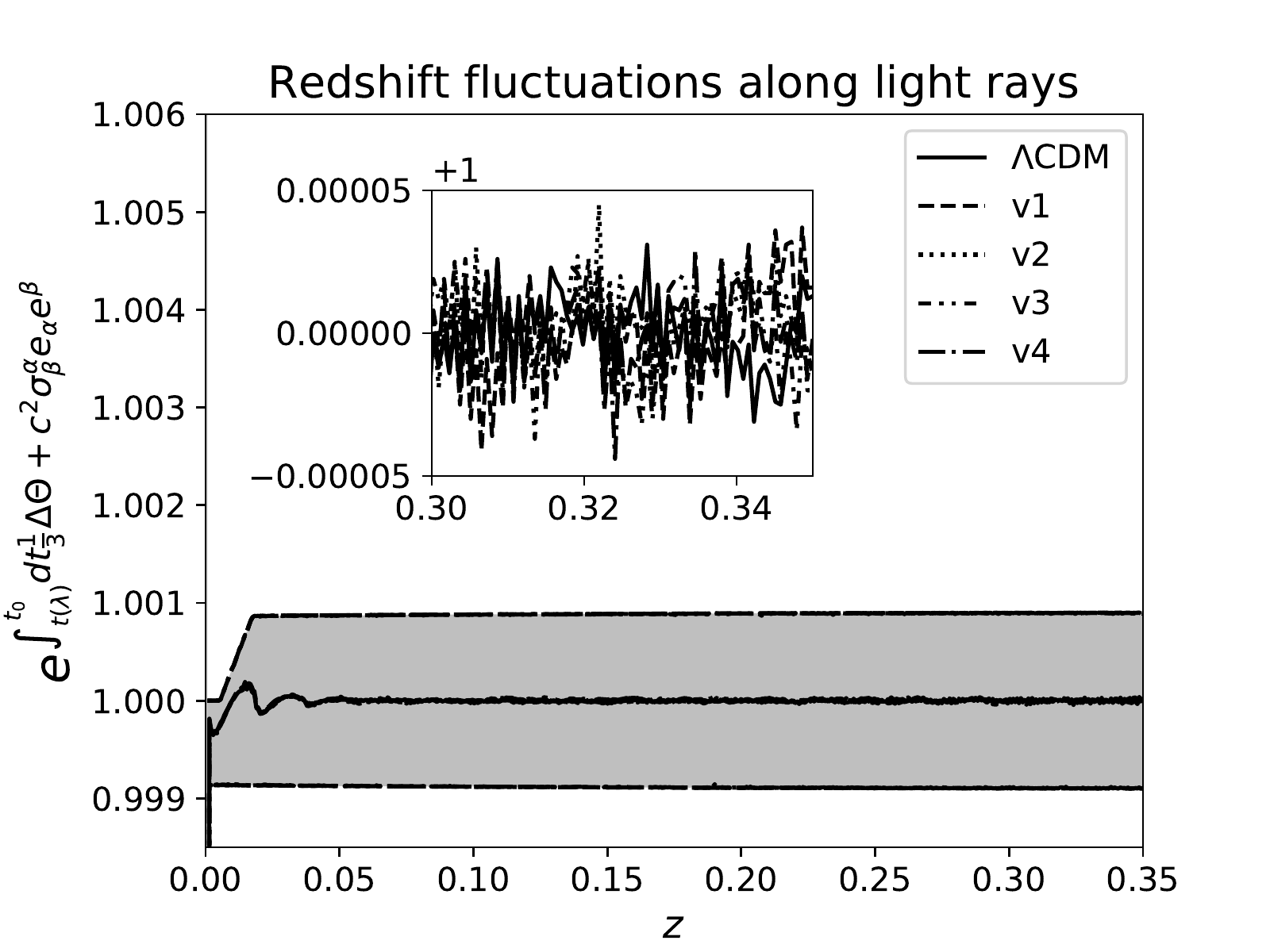}
\caption{Mean and dispersion of fluctuations in the redshift along light rays with five different realizations of 1000 random light rays. The shaded area indicates the dispersion for light rays with the same realization as in the main text. Close-ups of the mean values are included in the interval $z\in[0.3,0.35]$.}
\label{fig:sum_var}
\end{figure}

\begin{figure}
\centering
\includegraphics[scale = 0.5]{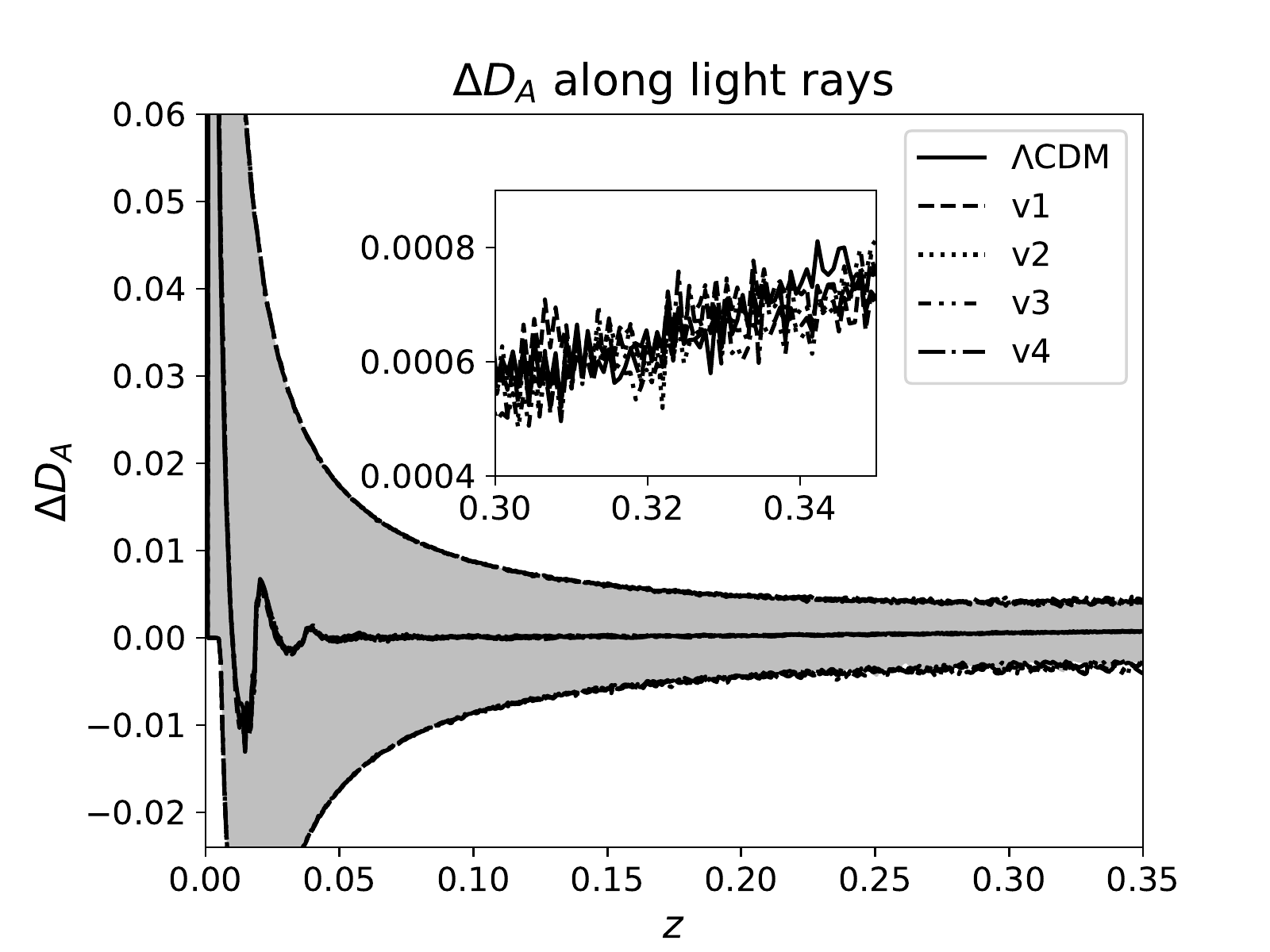}
\caption{Mean and dispersion of $\Delta D_A$ along light rays with five realizations of 1000 random light rays. The shaded area indicates the dispersion amongst light rays with the same realization as that used in the main text. Close-ups of the mean values are included in the interval $z\in[0.3,0.35]$.}
\label{fig:DA_var}
\end{figure}

\begin{figure*}
\centering
\subfigure[]{
\includegraphics[scale = 0.5]{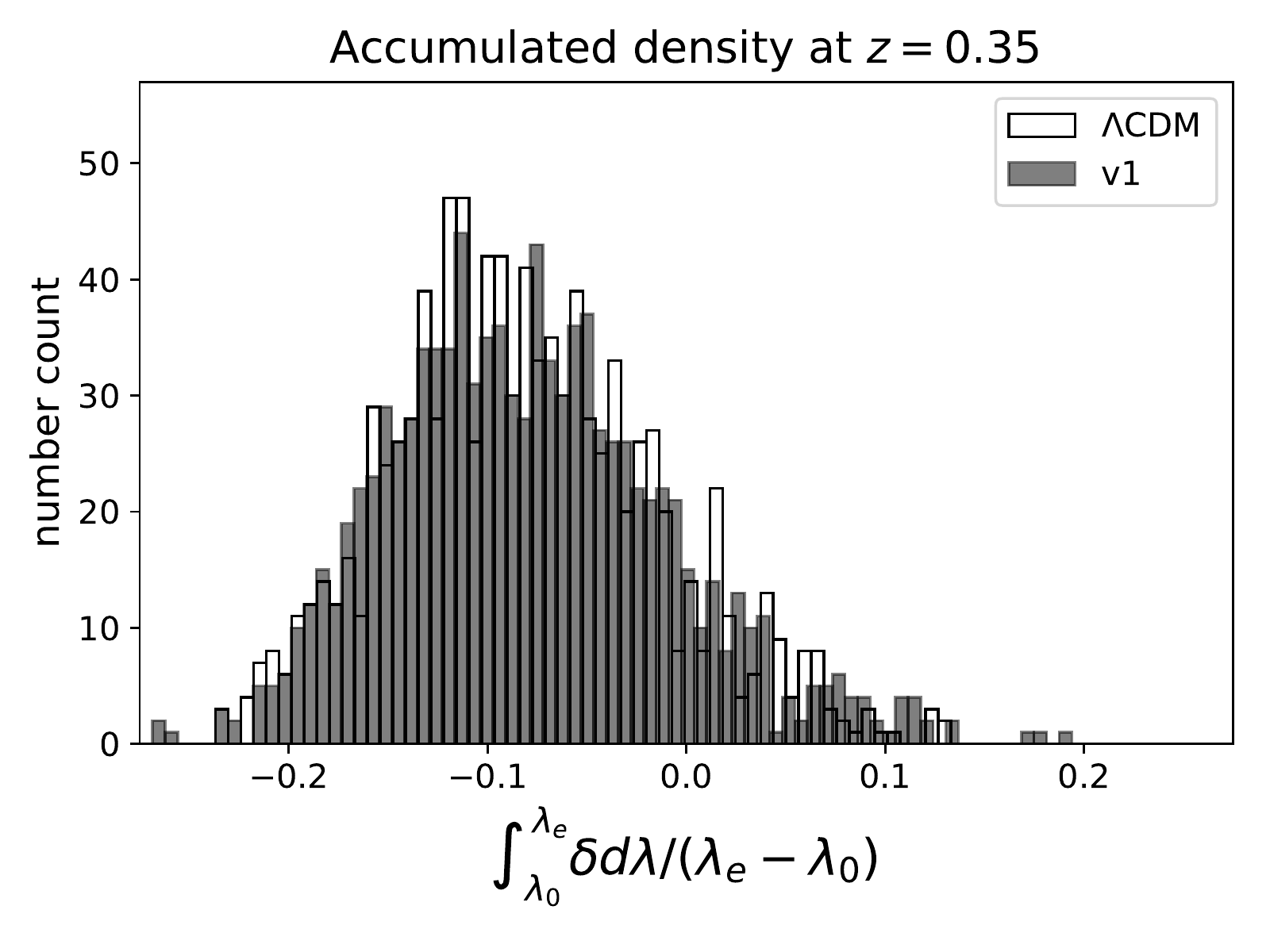}
}
\subfigure[]{
\includegraphics[scale = 0.5]{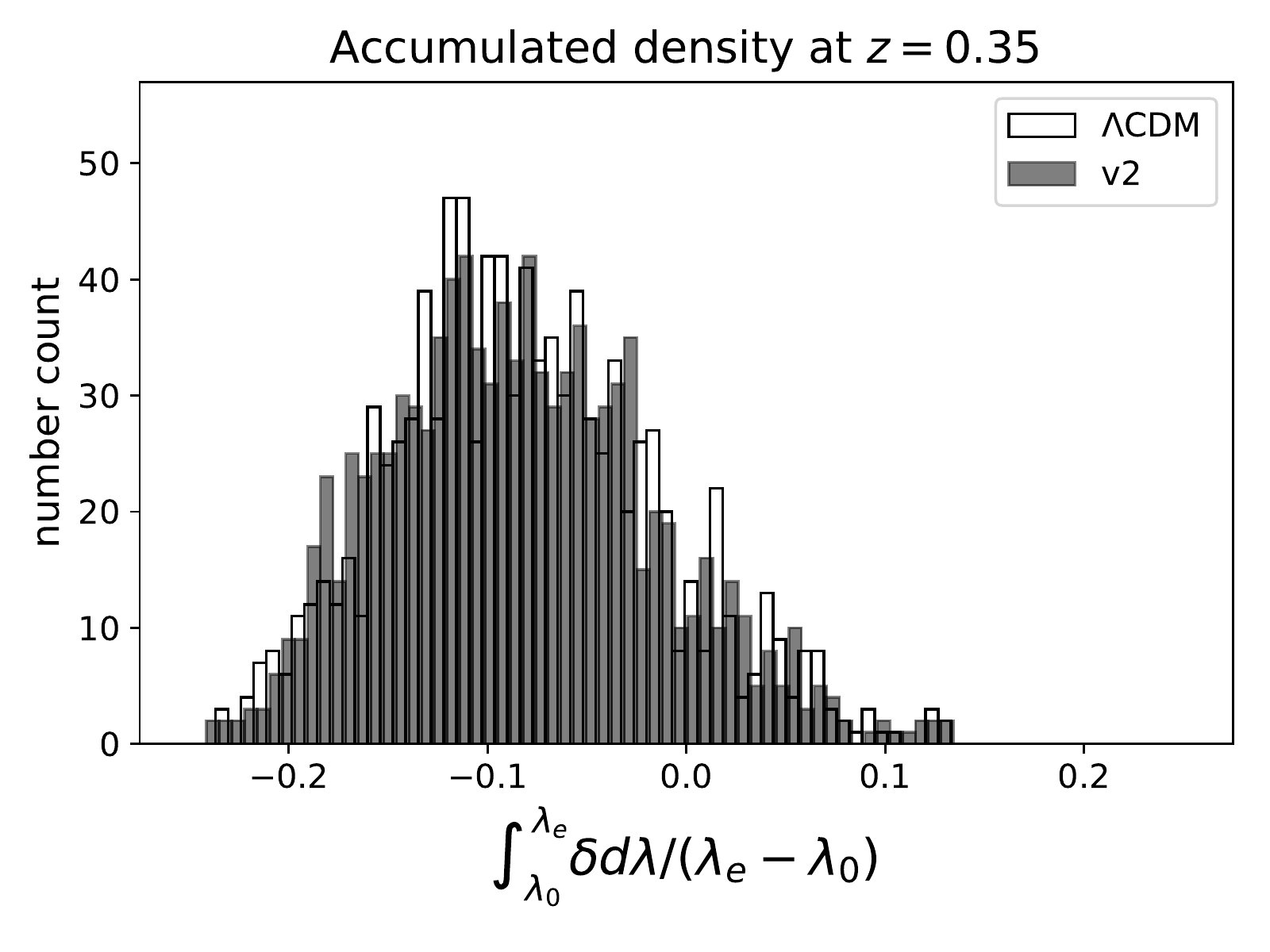}
}\par
\subfigure[]{
\includegraphics[scale = 0.5]{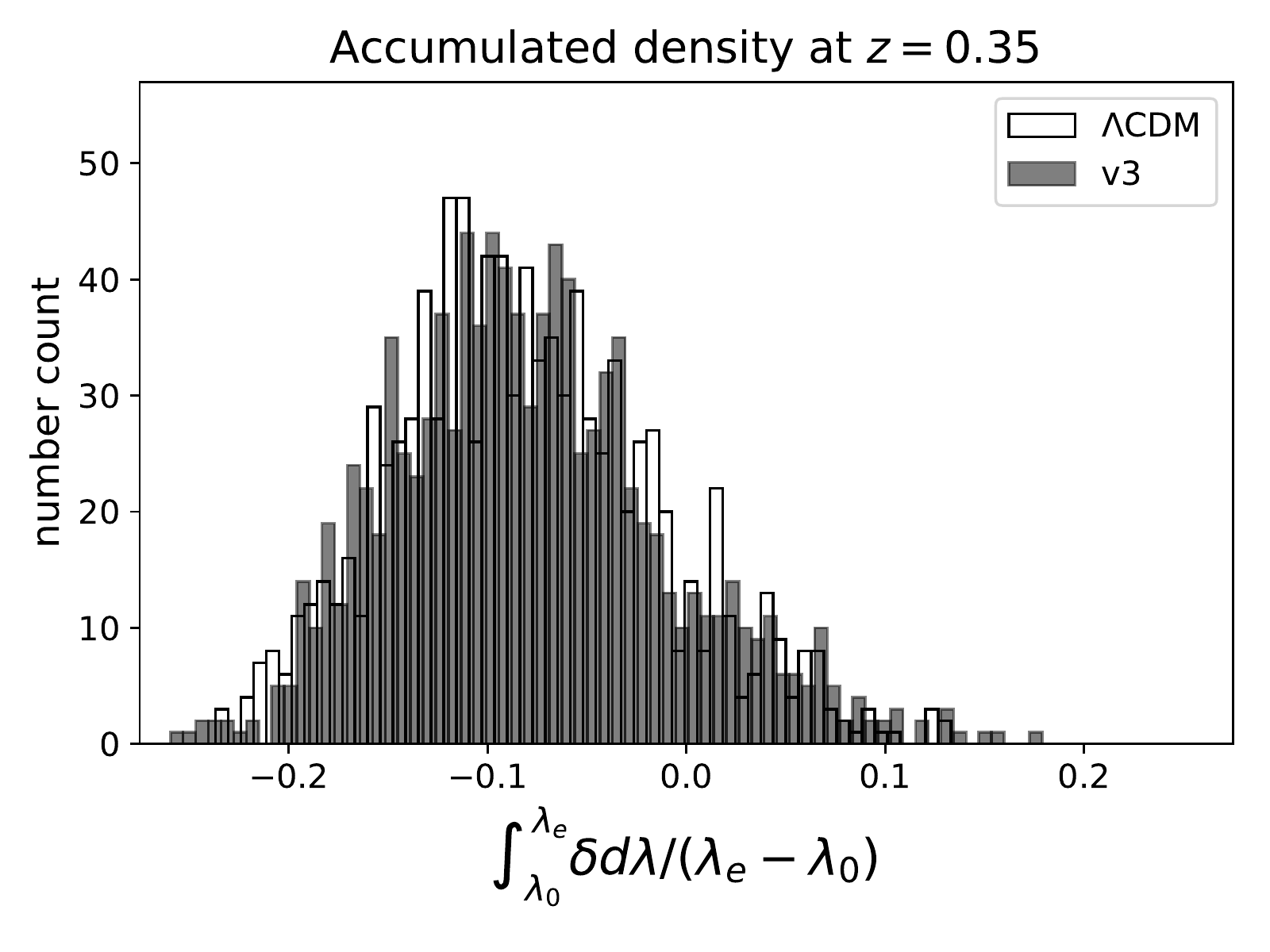}
}
\subfigure[]{
\includegraphics[scale = 0.5]{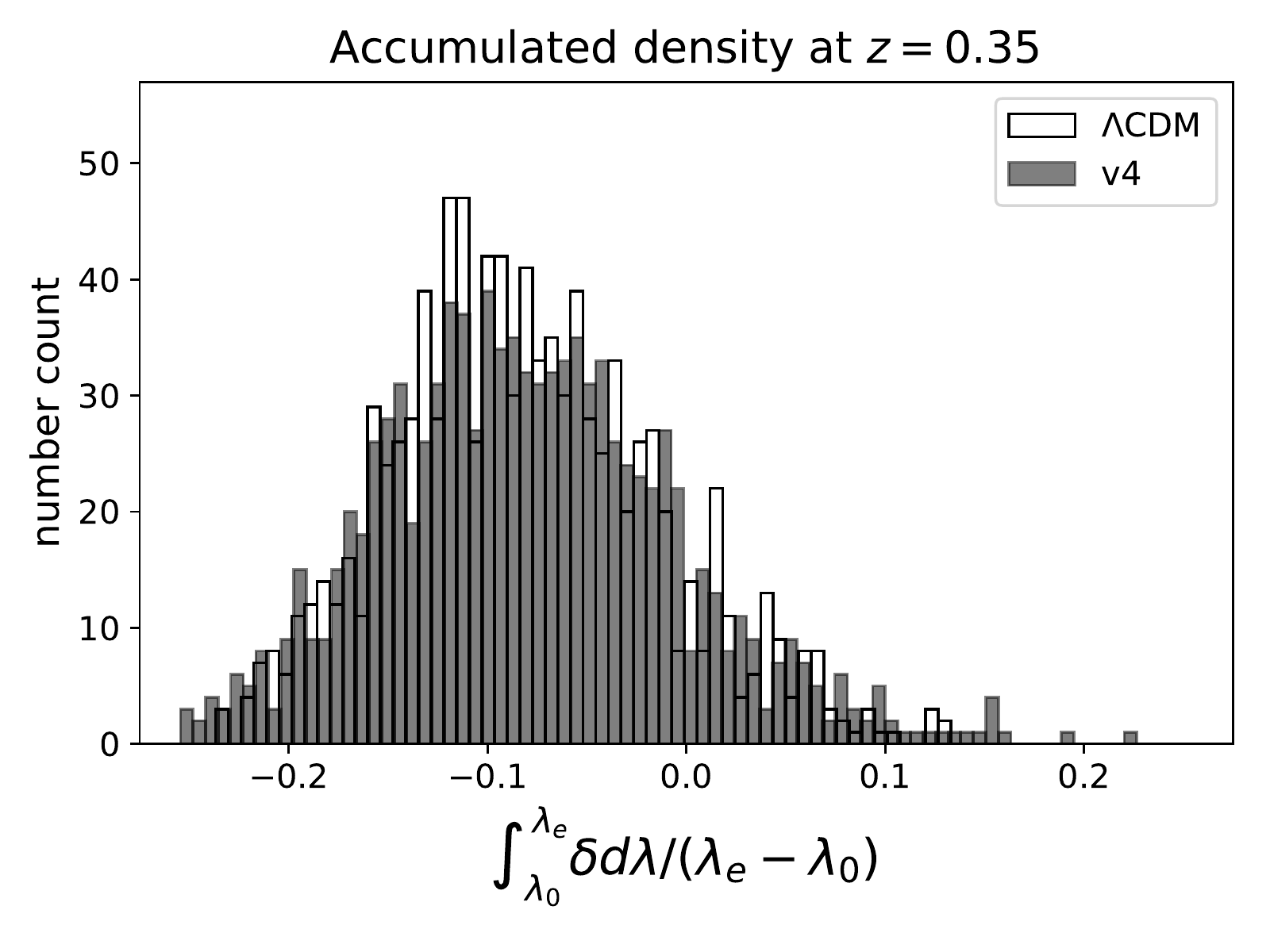}
}
\caption{Histograms showing the accumulated density contrast at $z = 0.35$ for five different realizations of 1000 random light rays. In each subfigure, results from the realization used in the main text is compared with results obtained with one of the other realizations. The particular realization that was used in the main text is labeled as $\Lambda$CDM while the other realizations are labeled as v1, v2, v3 and v4. Bin widths are approximately $0.006$. To ease comparison of the individual subfigures, these have all been given the same axis intervals as each other and as in the equivalent figure in the main text.}
\label{fig:rho_hist_var}
\end{figure*}

\begin{figure*}
\centering
\subfigure[]{
\includegraphics[scale = 0.5]{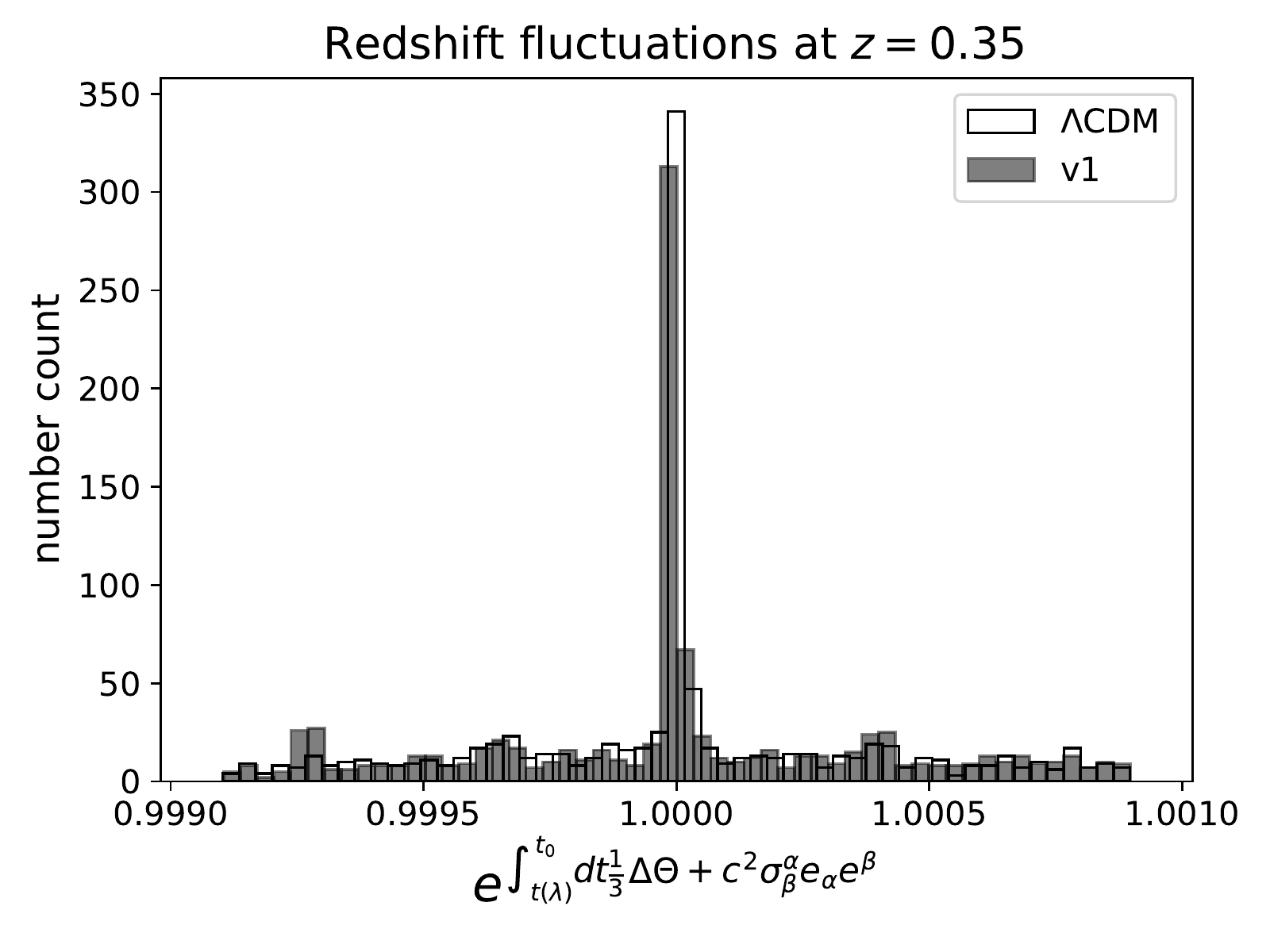}
}
\subfigure[]{
\includegraphics[scale = 0.5]{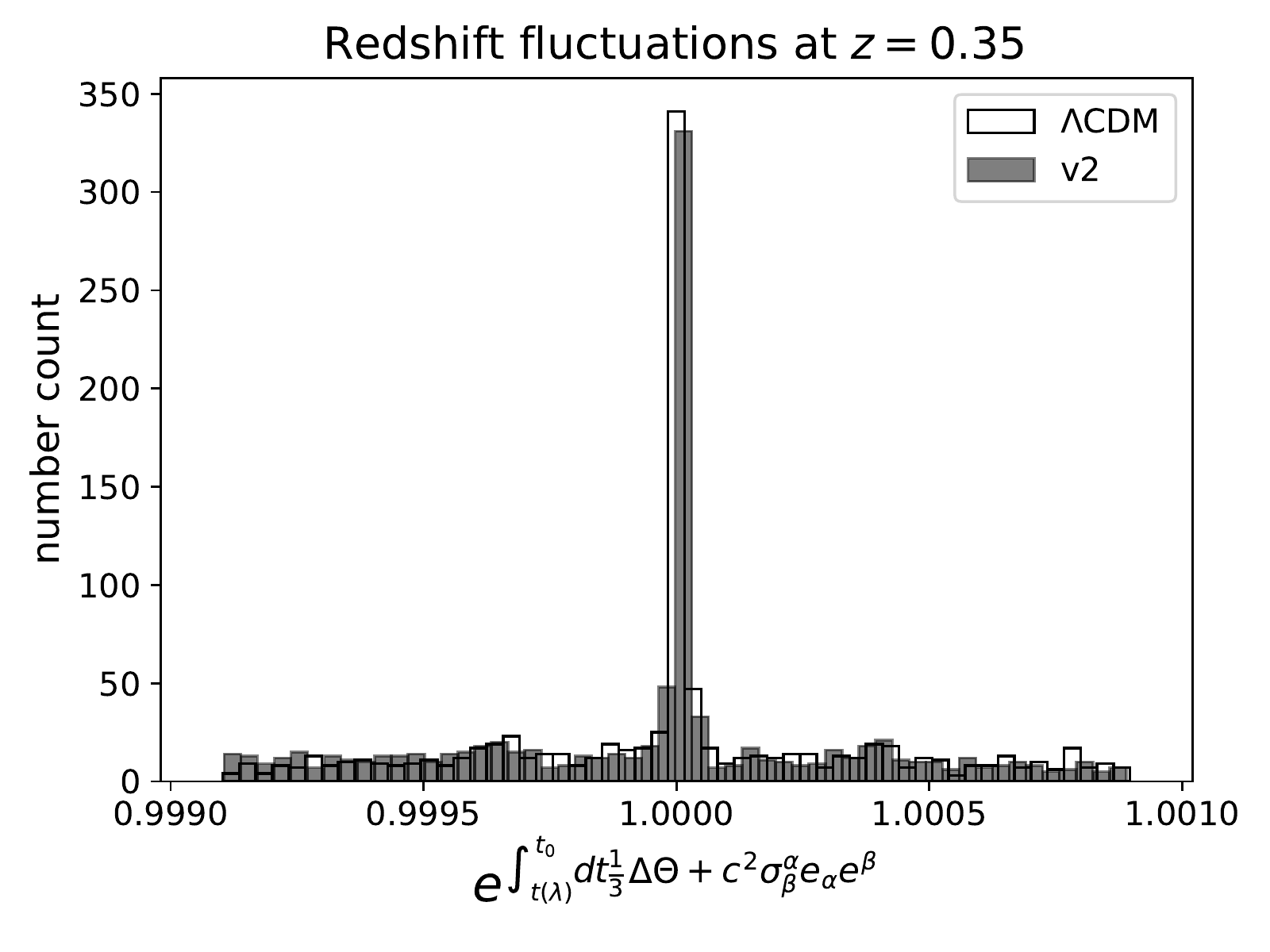}
}\par
\subfigure[]{
\includegraphics[scale = 0.5]{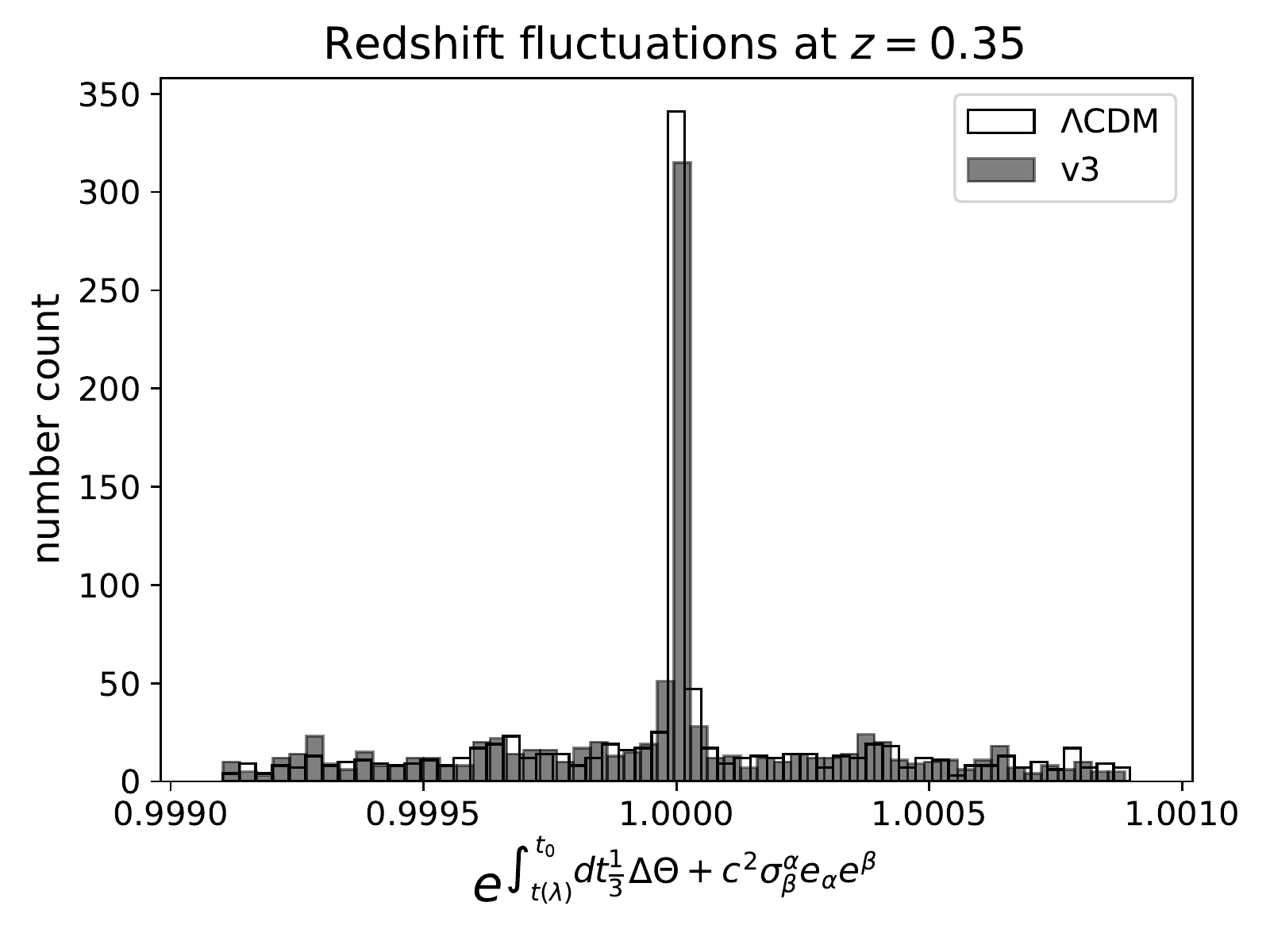}
}
\subfigure[]{
\includegraphics[scale = 0.5]{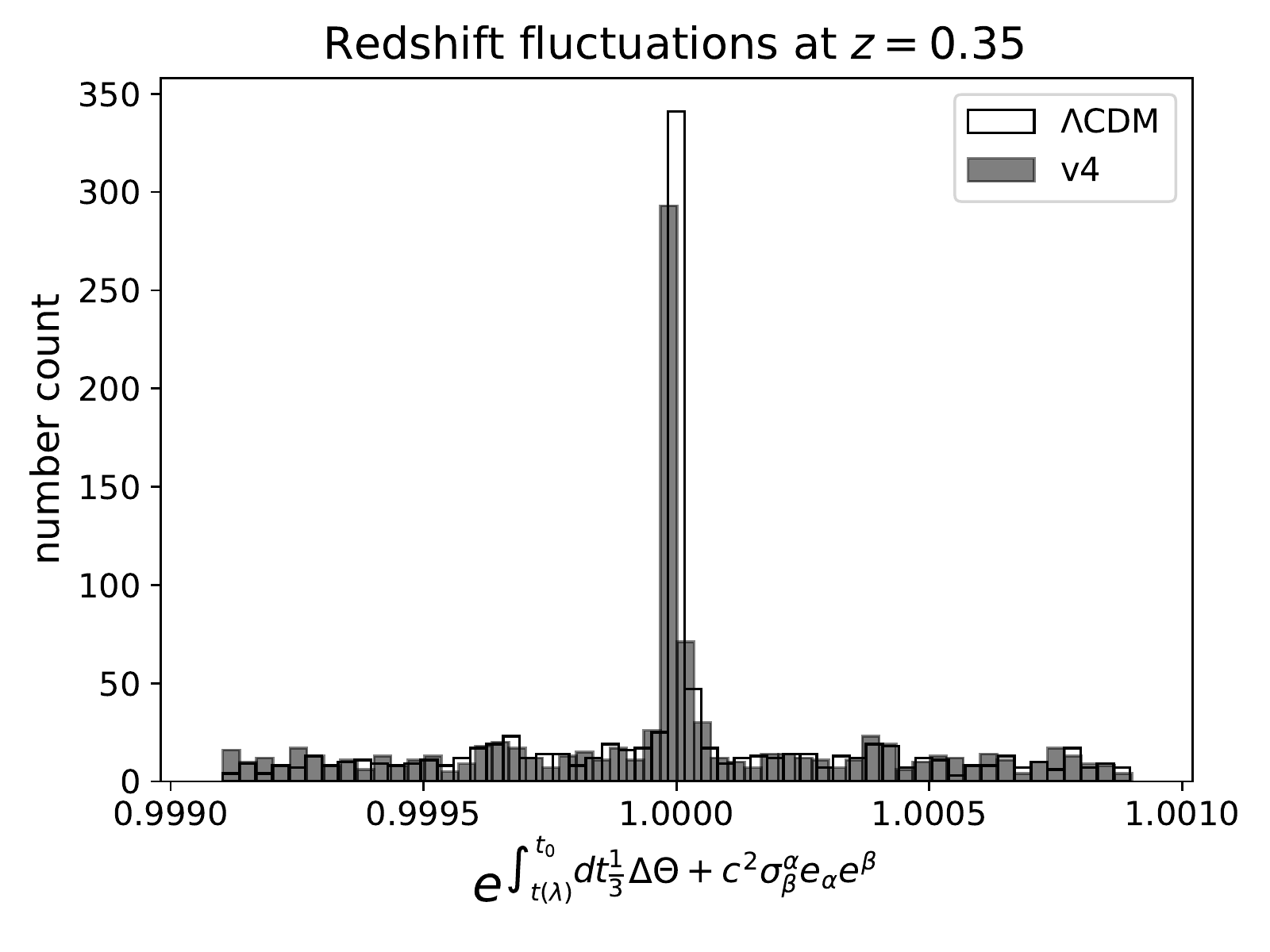}
}
\caption{Histograms showing the redshift fluctuations at $z = 0.35$ for five realizations of 1000 random light rays. In each subfigure, results from the model with the same realization as in the main text is compared with results based one of the other realizations. The particular realization that was used in the main text is labeled as $\Lambda$CDM while the other realizations are labeled as v1, v2, v3 and v4. Bin widths are approximately $0.000033$. To ease comparison of the individual subfigures, these have all been given the same axis intervals as each other and as in the equivalent figure in the main text.}
\label{fig:sum_hist_var}
\end{figure*}

\begin{figure*}
\centering
\subfigure[]{
\includegraphics[scale = 0.5]{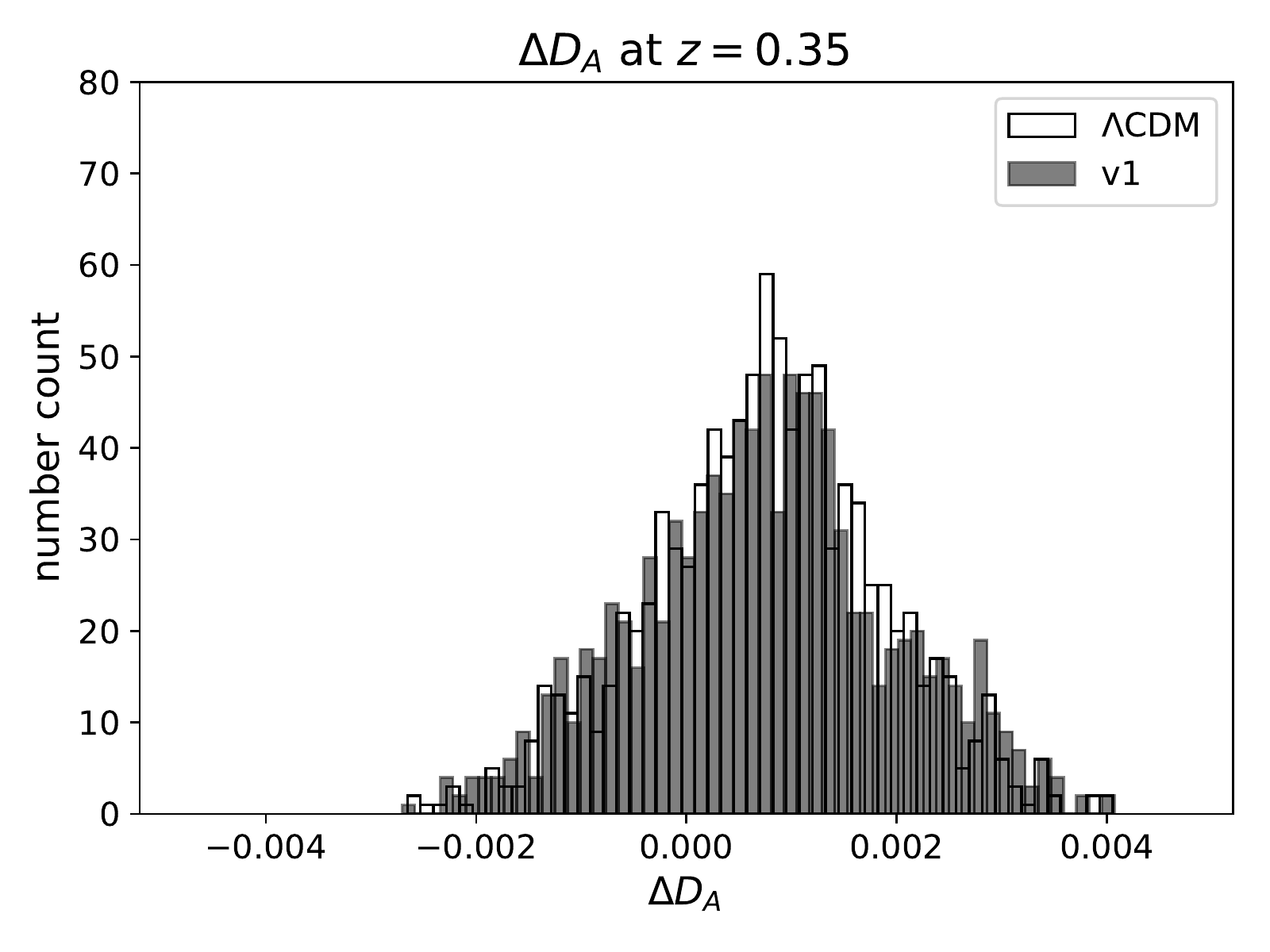}
}
\subfigure[]{
\includegraphics[scale = 0.5]{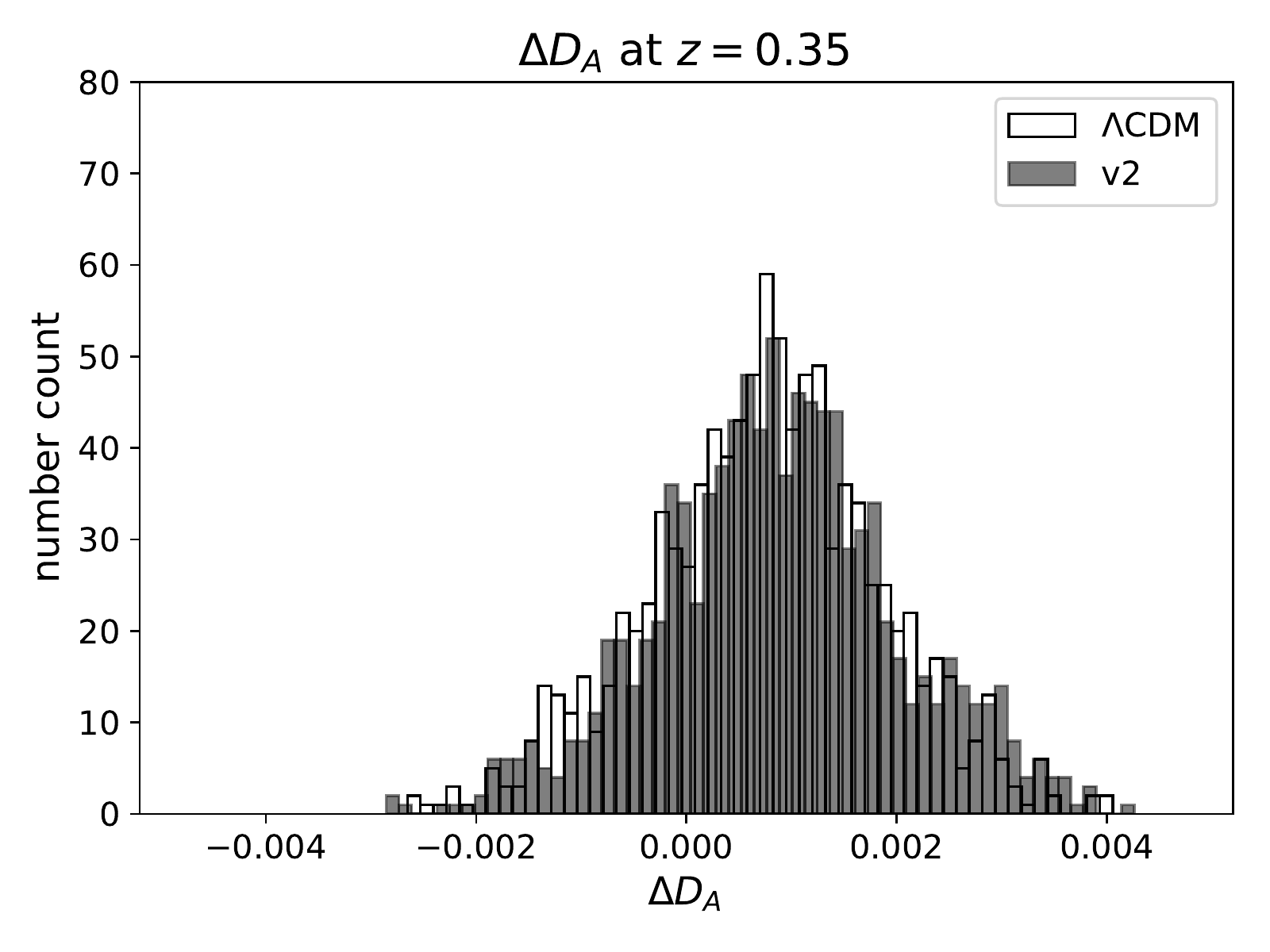}
}\par
\subfigure[]{
\includegraphics[scale = 0.5]{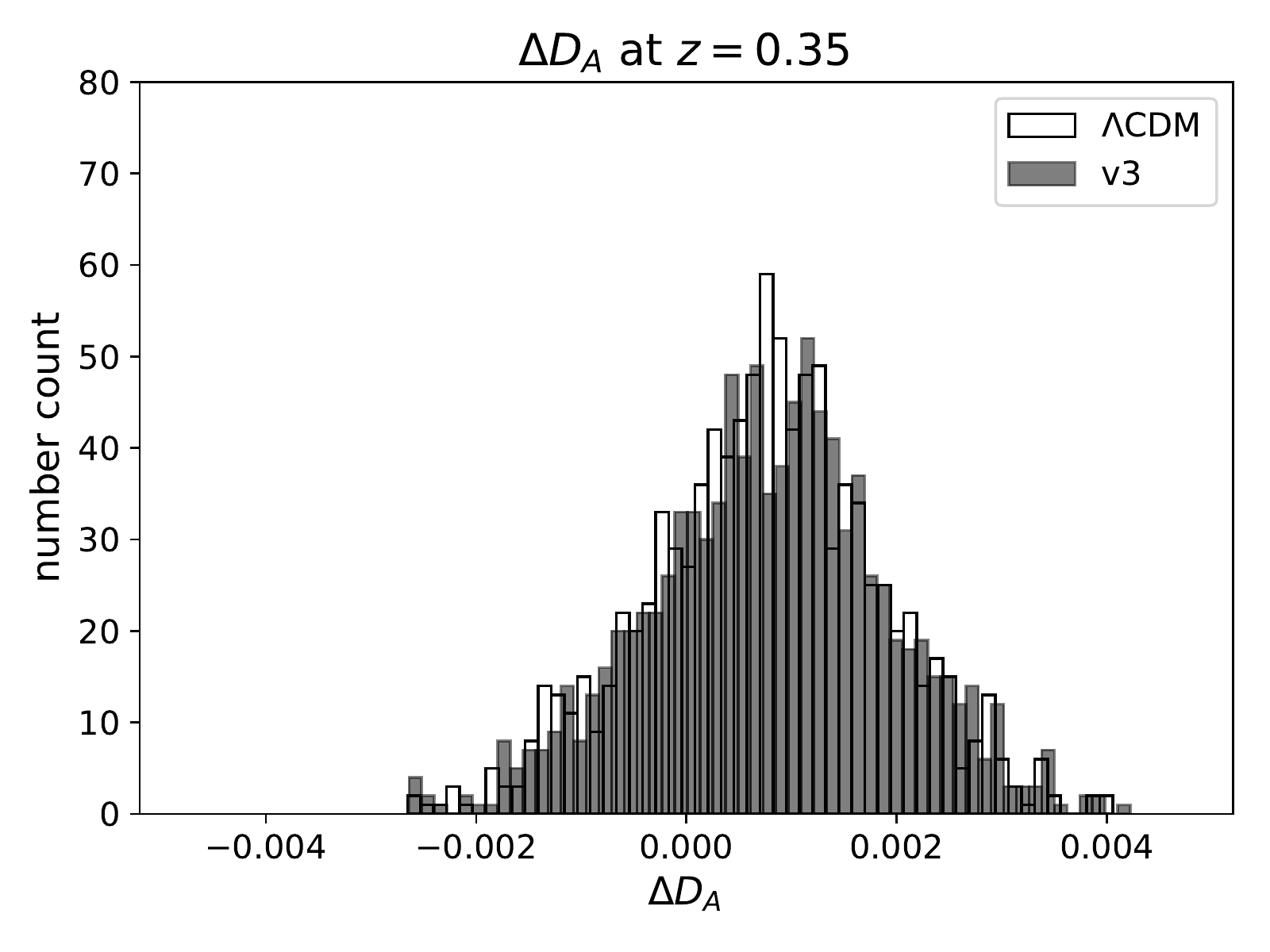}
}
\subfigure[]{
\includegraphics[scale = 0.5]{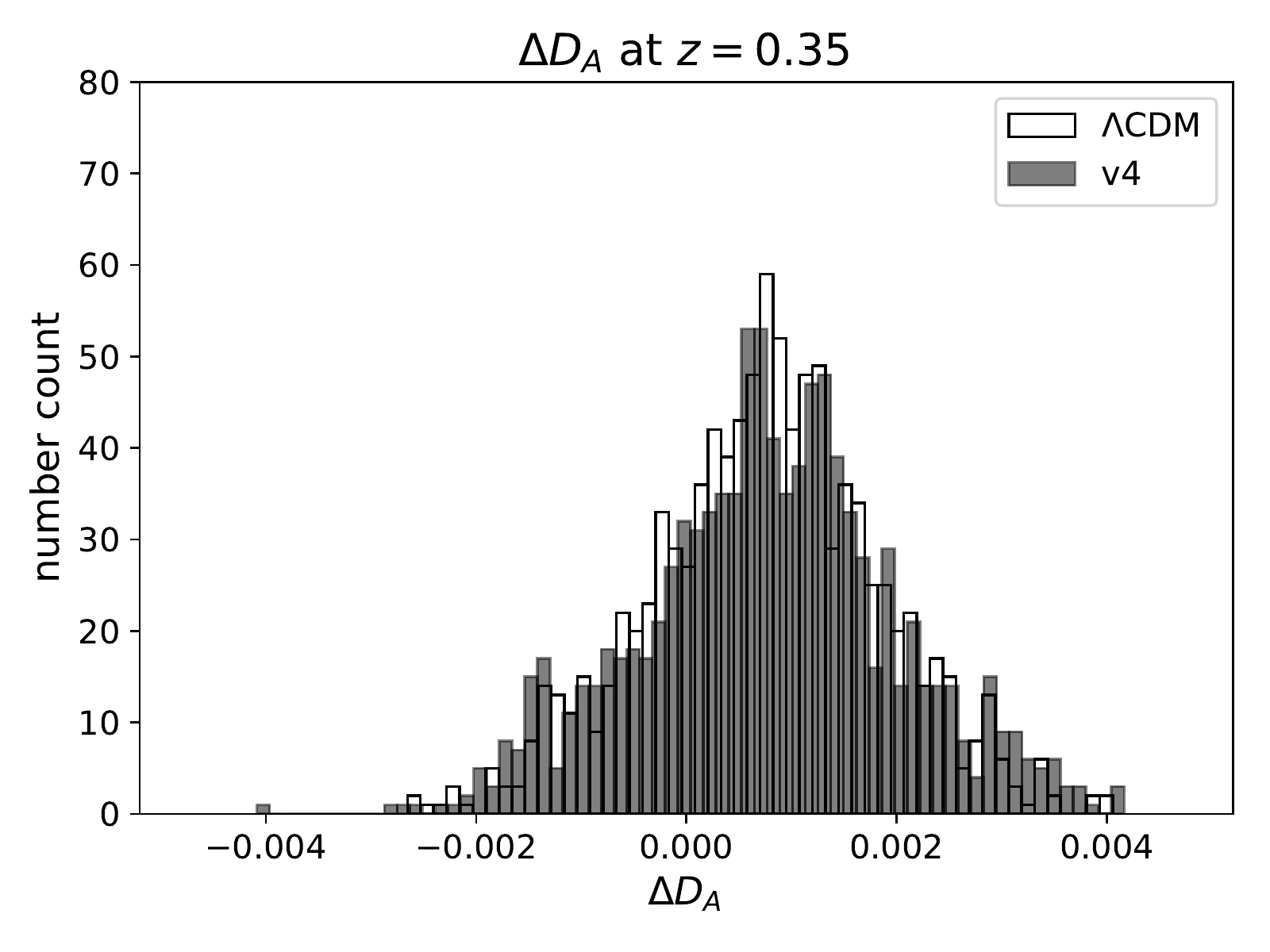}
}
\caption{Histograms showing fluctuations in $D_A$ at $z = 0.35$ for five realizations of 1000 random light rays in different Swiss cheese models. In each subfigure, results from the realization used in the main text is compared with results from one of the other realizations. The particular realization that was used in the main text is labeled as $\Lambda$CDM while the other realizations are labeled as v1, v2, v3 and v4. Bin widths are approximately $0.006$. To ease comparison of the individual subfigures, these have all been given the same axis intervals as each other and as in the equivalent figure in the main text.}
\label{fig:DA_hist_var}
\end{figure*}

Any possible sample variance should be irrelevant for the purpose of the study in the main text. Estimating its impact on the presented mean values and distributions is nonetheless enlightening and interesting in its own right and is helpful for understanding the results in the main text. This appendix therefore serves to study the significance of sample variance for the study in the main text where only 1000 light rays were studied per model. In order to asses this significance, the study has been replicated four times for the model with the standard $\Lambda$CDM background. The results from the five obtained data sets with the $\Lambda$CDM background are compared in the following.
\newline\newline
Figure \ref{fig:rho_var} shows the accumulated density contrast along 1000 light rays for the five different realizations of these. As seen from especially the close-up, the mean values are almost identical in each model at the higher end of the considered redshift interval, indicating that sample variance has low impact on the mean values at these redshifts. It is curious though, that the accumulated density contrast is negative since a light ray can sample an LTB model's overdensity without also sampling its underdensity, but not vice versa. The sign of this mean value could be due to a lack of rare-events light rays that sample mainly overdensities but which are so rare that they have not been sampled (sufficiently) by the 5000 light rays.
\newline\newline
Figure \ref{fig:sum_var} shows the fluctuations of the redshift compared to its background value. The fluctuations are shown for five different realizations of 1000 light rays and as seen, the results obtained from each model are very similar. Especially the mean values do not deviate much from each other. A similar result is seen in figure \ref{fig:DA_var} which shows the fluctuations in the angular diameter distance along the same light rays. One may note that the mean shift in the angular diameter distance is positive. This is in agreement with the general expectation that the most significant contribution to $\Delta D_A$ is the gravitational convergence which is given by the negative of an integral over the weighted density contrast along the given light ray. The sign of the mean values of $\Delta D_A$ found here are thus in agreement with the sign of the mean accumulated density contrasts. Note however, that such a comparison is not entirely accurate due to the differences in the weights on the density contrast in the integral of the gravitational convergence and the accumulated density contrast computed here. In addition, it was recently shown in \cite{dig_og_cc} that e.g. the lowest order Born correction to $\Delta D_A$ can become numerically larger than the gravitational convergence along some lines of sight. This contribution is second order in perturbation theory though, so if one averages over ``enough'' light rays, this contribution should vanish according to \cite{angle_vs_ensemble}. Note lastly that the sign of the mean value of $\Delta D_A$ found here is in agreement with what is expected for ensemble averages based on second order perturbation theory \cite{angle_vs_ensemble}.
\newline\newline
In addition to studying the impact of sample variance on the mean and dispersion of the accumulated density contrast, redshift fluctuations and fluctuations in the redshift-distance relation, the impact of sample variance on the actual distributions of these quantities has also been studied. Specifically, histograms as those in the main text are here shown for the different realizations of 1000 light rays in the model with the $\Lambda$CDM background.
\newline\indent
In figure \ref{fig:rho_hist_var}, the distribution of the accumulated density contrast at $z = 0.35$ is shown. The figure contains four subfigures, each comparing the distribution of a given realization with that of the realization used in the main text. Although the five distributions are quite similar, they show noticeable differences such as some having small positive tails with maximum value in the accumulated density contrast almost twice as large in some realizations compared to some of the other realizations. A similar situation was found in the main text where it was suggested to be the result of different density contrasts of the studied models. Here, only a single model is studied and hence the only explanation is sample variance. The reason for having positive tails in the accumulated density contrast without any similar negative tails must be due to the fact that, as mentioned above, a light rays can sample the overdensity of a structure without sampling it underdensity while the opposite is not possible for the specific Swiss cheese models studied here.
\newline\indent
Note that it does not constitute an inconsistency that the tails are here attributed variance while they in the main text are attributed the size of density contrasts. Indeed, the specific setups in the main text and in this appendix each exclude the explanation given in the other part of the text. As such, the tails found here would be expected to be even greater for the particular realizations of 1000 light rays in the models that already for the realization used in the main text exhibit tails.
\newline\newline
Figure \ref{fig:sum_hist_var} shows the distributions of the fluctuations (compared to the background value) of the redshift at $z = 0.35$. The differences between the different realizations' distributions are quite small. It is especially noticeable that the maximum and minimum values of the fluctuations are nearly identical in each realization. A similar result is found in figure \ref{fig:DA_hist_var} which shows the distribution in the angular diameter distance at $z = 0.35$. This could be an indication that the main quantity responsible for these dispersions is the background expansion rate since this quantity is the same for the studied realizations. This suggestion fits well with the equivalent results shown in the main text.
\newline\newline
Lastly, it may be noted that the similarities in the mean values and the distributions about the means indicate a relatively high statistical significance of the results. This is somewhat surprising considering that very low statistical significances were obtained in both \cite{cmb_digselv,syksyCMB} despite those studies being based on a much larger number of light rays. A relatively large statistical significance was also found in \cite{Ishak} which also only considered 1000 light rays (in Swiss cheese models based on the quasi-spherical Szekeres model). As noted in \cite{cmb_digselv} this seeming inconsistency could be due to the much larger packing fractions obtained when constructing Swiss cheese models on the fly (as here and in \cite{Ishak}) compared to the maximum packing fraction of $\sim 0.64$ obtainable for fixed Swiss cheese models with a random distribution of structures of a single size (see e.g. \cite{RCP_experimental}).

\FloatBarrier

\end{document}